\documentstyle[12pt,epsf]{article}

\advance \topmargin by -\headheight
\advance \topmargin by -\headsep

\evensidemargin \oddsidemargin
\def\ie{{\em i.e.}}

\def\ie{\hbox{\it i.e.}}

\def\CC{{\mathchoice
{\rm C\mkern-8mu\vrule height1.45ex depth-.05ex 
width.05em\mkern9mu\kern-.05em}
{\rm C\mkern-8mu\vrule height1.45ex depth-.05ex 
width.05em\mkern9mu\kern-.05em}
{\rm C\mkern-8mu\vrule height1ex depth-.07ex 
width.035em\mkern9mu\kern-.035em}
{\rm C\mkern-8mu\vrule height.65ex depth-.1ex 
width.025em\mkern8mu\kern-.025em}}}

\def\RR{{\rm I\kern-1.6pt {\rm R}}}

\def\ZZ{{\rm Z}\kern-3.8pt {\rm Z} \kern2pt}

\def\np{Nucl. Phys.}
\def\pl{Phys. Lett.}
\def\prl{Phys. Rev. Lett.}
\def\pr{Phys. Rev.}

\def\sjnp{Sov. J. Nucl. Phys.}

\def\atmp{Adv. Theor. Math. Phys. }
\def\jhep{J. High Energy Phys.}
\def\ptp{Prog. Theor. Phys.}

\newcommand{\beq}{\begin{equation}}
\newcommand{\eeq}{\end{equation}}
\newcommand{\rc}{\nonumber\\}
\newcommand{\bear}{\begin{eqnarray}}
\newcommand{\eear}{\end{eqnarray}}

\newfont{\namefont}{cmr10}
\newfont{\addfont}{cmti7 scaled 1440}
\newfont{\boldmathfont}{cmbx10}
\newfont{\headfontb}{cmbx10 scaled 1728}
\renewcommand{\theequation}{{\rm\thesection.\arabic{equation}}}
\begin{document}
\begin{titlepage}

\begin{center} \Large \bf Worldvolume Dynamics of D-branes in a D-brane
Background

\end{center}

\vskip 0.3truein
\begin{center} 
J. M. Camino
\footnote{e-mail:camino@fpaxp1.usc.es}, 
A.V. Ramallo
\footnote{e-mail:alfonso@fpaxp1.usc.es}
and 
J. M. S\'anchez de Santos 
\footnote{e-mail:santos@gaes.usc.es}

\vspace{0.3in}

Departamento de F\'\i sica de
Part\'\i culas, \\ Universidad de Santiago\\
E-15706 Santiago de Compostela, Spain. 
\vspace{0.3in}

\end{center}
\vskip 1truein

\begin{center}
\bf ABSTRACT
\end{center} 

We study the embedding of D(8-p)-branes in the background geometry of
parallel \break Dp-branes for $p\le 6$. The D(8-p)-brane is extended
along the directions orthogonal to the Dp-branes of the background. 
The D(8-p)-brane configuration is determined by its Dirac-Born-Infeld
plus Wess-Zumino-Witten action. We find a BPS condition which solves
the equation of motion. The analytical solution of the BPS equation
for the near-horizon and asymptotically flat geometries  is given. The
embeddings we obtain represent branes joined by tubes. By analyzing the
energy of these tubes we conclude that they can be regarded as bundles
of fundamental strings. Our solution provides an explicit realization
of the Hanany-Witten effect. When $p=6$ the solution of the BPS
equation  must be considered separately and, in general, the
embeddings of the D2-branes do not admit the same interpretation as in
the $p<6$ case.

\vskip4.5truecm
\leftline{US-FT-12/99 \hfill May 1999}
\leftline{hep-th/9905118}
\smallskip
\end{titlepage}
\setcounter{footnote}{0}

%%%%%%%%%%%%%%%%%%%%%%%%%%%%%%%%%%%%%%%%%%%%%%%%%%%%%%%%%%%%%%%%%%%%%%%%%%%%%%
%%%%%          		    M A I N   T E X T 
%%%%%%%%%%%%%%%%%%%%%%%%%%%%%%%%%%%%%%%%%%%%%%%%%%%%%%%%%%%%%%%%%%%%%%%%%%%%%%

\setcounter{equation}{0}
\section{Introduction}
\medskip

D-branes have been shown to play a fundamental role in the
non-perturbative structure of string theory. Originally
\cite{Pol}, they were
introduced as hypersurfaces in space-time on which open strings with
Dirichlet boundary conditions are allowed to end. The  extended
objects so defined are charged under the Ramond-Ramond (RR) sector of
the theory and are essential to understand the duality structure of
string theories. Moreover, the low energy physics of D-branes can be
described by supersymmetric gauge theory and, actually, super
Yang-Mills theory arises as a low energy description of parallel
D-branes.

D-branes also appear as solutions of the supergravity effective action
of string theory \cite{supergravity}. These solutions are extended
generalizations of the Schwarzschild geometry which describe the
gravitational field created by an object carrying Ramond-Ramond charge.
This gravitational field acts on the dimensions orthogonal to the
D-brane worldvolume which, from the point of view of the gauge theory,
can be  regarded as ``extra dimensions". 

The interplay between gauge theory and gravity is on the basis of
the Maldacena conjecture \cite{Maldacena}. Indeed, Maldacena has argued
that there is a remarkable duality between classical supergravity in the
near-horizon region and the large $N$ 't Hooft limit of the
$SU(N)$ Yang-Mills theory. In particular, the near-horizon geometry of
parallel D3-branes is equivalent to the space $AdS_5\times S^5$, whose
boundary can be identified with four-dimensional Minkowsky space-time.
In this case one gets the so-called $AdS/CFT$ correspondence between
type IIB superstrings  on $AdS_5\times S^5$ and ${\cal N}=4$
non-abelian super Yang-Mills theory in four dimensions \cite{Gubser,
Witten, Gross}.  Within this context, one can
compute Wilson loop expectation values by considering a fundamental
open string placed on the interior of the $AdS_5$ space and having its
ends on the boundary \cite{Wloop}. This formalism has been used to
extract quark potentials both in the supersymmetric and
non-supersymmetric theory. Moreover, Witten \cite{Wittenbaryon} has
proposed a way to incorporate baryons by means of  a D5-brane wrapped
on the   $S^5$, from which  fundamental strings that end on the
D3-branes emanate \cite{Branbaryon, Imabaryon}. 

In ref. \cite{CGS1}, the problem of a D5-brane moving under the
influence of a D3-brane background was considered. The D5-brane
dynamics is governed by an action which is the sum of a
Dirac-Born-Infeld and a Wess-Zumino term. The equations of motion of
the static \mbox{D5-brane} are solved if one assumes that the fields
satisfy a certain first-order BPS differential equation that was found
previously by Imamura \cite{Imamura}. This BPS  condition can be
integrated exactly in the near-horizon region of the background
D3-brane geometry and the solutions are spikes of the D5-brane
worldvolume which can be interpreted as bundles of fundamental strings
ending on the D3-branes. Actually, these kinds of spikes are a general
characteristic of the Dirac-Born-Infeld non-linear gauge theory and can
be used to describe strings attached to branes \cite{CM, Gibbons}. 

The BPS condition for the full asymptotically flat D3-brane metric was
analyzed numerically in ref. \cite{CGS1}. As a result of this study one
gets a precise description of the string creation process which takes
place when two D-branes cross each other, \ie\ of  the so-called
Hanany-Witten effect \cite {HW, HWothers}. These results were
generalized in ref. \cite{CGS2} to the case of baryons in confining
gauge theories. Moreover, in ref. \cite{Craps} the BPS differential
equation was obtained as a condition which must be fulfilled in order
to saturate  a lower bound on the energy.

In this paper we shall study the worldvolume dynamics of a
D(8-p)-brane  moving under the action of the gravitational and
Ramond-Ramond fields of a stack of background \mbox{Dp-branes} 
for $p\le 6$.  The D(8-p)-brane is extended along the directions
orthogonal to the worldvolume of the background Dp-branes. For $p=3$
this system is the one analyzed in ref. \cite{CGS1}. By using the
method of ref.  \cite{Craps}, we
will find a BPS first-order differential equation whose solutions also
verify the equation of motion. Remarkably, we will be able to
integrate analytically this BPS equation both in the near-horizon and
asymptotically flat metrics. For $p\le 5$ the solutions we
will find are similar to the ones described in ref. \cite{CGS1}, \ie\
the  D(8-p)-brane worldvolume has spikes which can be interpreted as
flux tubes made up of fundamental strings. Our analytical solution will
allow us to give a detailed description of the shape of these tubes,
their energy and of the process of string creation  from D-brane
worldvolumes. 

The organization of this paper is the following. In section 2 we
formulate our problem. The BPS differential equation is obtained in
section 3. In section 4 this BPS condition is integrated for $p\le 5$
in the near-horizon region. The $p=6$ system requires a special
treatment, which is discussed in section 5. The solution of the BPS
equation for the full \mbox{Dp-brane} geometry is obtained in section 6.
The detailed analysis of this solution is performed in appendix A.
Finally, in section 7, we summarize our results and point out some
directions for future work.

\setcounter{equation}{0}
\section{Worldvolume  Dynamics}
\medskip
The ten-dimensional metric (in the string frame) corresponding to a
stack of $N$ coincident extremal Dp-branes in type II superstring theory
is \cite{supergravity}:
\beq
ds^2\,=\,\Bigl[\,H_p(r)\,\Bigr]^{-{1\over 2}}\,\,
(\,-dt^2\,+\,dx_{\parallel}^2\,)\,+\,
\Bigl[\,H_p(r)\,\Bigr]^{{1\over 2}}\,\,
(\,dr^2\,+\,r^2\,d\Omega_{8-p}^2\,)\,\,.
\label{uno}
\eeq
In eq.  (\ref{uno}) $x_{\parallel}$ are $p$-dimensional cartesian
coordinates along the branes, $r$ is a radial coordinate
parametrizing the distance to the branes and $d\Omega_{8-p}^2$ is the
line element of an unit $8-p$ sphere. The harmonic function 
$H_p(r)$ appearing in eq. (\ref{uno}) is given by:
\beq
H_p(r)\,=\,a\,+\,\Bigl(\,{R\over r}\Bigr)^{7-p}\,\,,
\label{dos}
\eeq
where $a=0,1$ and the `radius' $R$ can be written in terms of the
number $N$ of branes, the Regge slope $\alpha\,'$ and the string
coupling constant $g_s$ as follows:
\beq
R^{7-p}\,=\,N\,g_s\,2^{5-p}\,\pi^{{5-p\over 2}}\,
(\,\alpha\,'\,)^{{7-p\over 2}}\,\,
\Gamma\Bigl(\,{7-p\over 2}\Bigr)\,\,.
\label{tres}
\eeq
We have chosen our conventions in such a way that the tension of the
fundamental string $T_{f}$ is: 
\beq
T_{f}\,=\,{1\over 2\pi\alpha\,'}\,\,.
\label{cuatro}
\eeq
The value $a=1$ in eq. (\ref{dos}) corresponds to the full geometry of
the stack of Dp-branes. Actually, in order to have an asymptotically
flat space-time, we shall restrict ourselves to the case $p<7$. Taking
$a=0$ in the harmonic function $H_p(r)$ is equivalent to approximate
the Dp-brane geometry by the near-horizon metric of the `throat'
region of (\ref{uno}). 

The Dp-brane solution we are considering is
also characterized by a dilaton field $\phi(r)$ and a 
Ramond-Ramond (RR) $p+2$-form field
$G$. The corresponding values are:
\bear
e^{-\tilde\phi(r)}\,&=&\,\Bigl[\,H_p(r)\,\Bigr]^{{p-3\over 4}}
\,\,,\rc\rc
G_{t\,,\,x_{\parallel}^1\,,\,\cdots\,\,,\,x_{\parallel}^p,\,r}\,&=&\,
{d\over dr}\,\Bigl[\,H_p(r)\,\Bigr]^{-1}\,\,,
\label{cinco}
\eear
where $\tilde\phi(r)\,=\,\phi(r)\,-\,\phi_{\infty}$ and
$\phi_{\infty}$ is the value of the dilaton field at infinite distance
of the Dp-branes 
(\ie\ $\phi_{\infty}\,=\,\lim_{r\rightarrow\infty}\,\phi(r)$). As was
pointed out in ref. \cite{Maldacena}, in order to trust this type II
supergravity solution, both the curvature in string units and the
dilaton must be small. This fact introduces restrictions in the values
of the radial coordinates for which the correspondence between the
supergravity and gauge theory descriptions is valid (see also 
ref.~\cite{charges}).

Let us now consider a D(8-p)-brane embedded in the transverse
directions of the stack of the background Dp-branes. The dynamics of
this  D(8-p)-brane is determined by its action, which is a sum of a
Dirac-Born-Infeld and a Wess-Zumino term:
\beq
S\,=\,-T_{8-p}\,\int d^{\,9-p}\xi\,e^{-\tilde\phi}\,
\sqrt{-{\rm det}\,(\,g\,+\,F\,)}\,+\,
T_{8-p}\,\int d^{\,9-p}\xi\,\, A\wedge\,^*G\,\,,
\label{seis}
\eeq
where $g$ is the induced metric on the worldvolume of the
D(8-p)-brane, $A$  is a worldvolume abelian gauge field and $F=dA$ its
field strength.  In eq. (\ref{seis}) $^*G$ is a (8-p)-form, which is
the pullback of the Hodge dual of the background RR (p+2)-form $G$.
Notice that this RR field acts as a source for the worldvolume gauge
field $A$. The coefficient $T_{8-p}$ appearing in the action
(\ref{seis}) is the tension of a  D(8-p)-brane, which is given by
\cite{Pol}:
\beq
T_{8-p}\,=\,(2\pi)^{p-8}\,(\,\alpha\,'\,)^{{p-9\over 2}}\,
(\,g_s\,)^{-1}\,\,.
\label{siete}
\eeq
Let $\theta^1$, $\theta^2$, $\cdots$, $\theta^{8-p}$ be coordinates
which parametrize the $S^{8-p}$ transverse sphere. The worldvolume
coordinates $\xi^{\alpha}$ ($\alpha\,=\,0\,,\,\cdots\,,\,8-p\,)$ will
be taken as:
\beq
\xi^{\alpha}\,=\,(\,t\,,\,\theta^1\,,\,\cdots\,,\,\theta^{8-p}\,)\,\,.
\label{ocho}
\eeq
We shall assume that the $\theta$'s are spherical angles on $S^{8-p}$
and that $\theta\equiv\theta^{8-p}$ is the polar angle 
($0\le\theta\le\pi$). Therefore, the $S^{8-p}$ line element 
$d\Omega_{8-p}^2$ can be decomposed as:
\beq
d\Omega_{8-p}^2\,=\,d\theta^2\,+\,(\,{\rm sin}\,\theta)^{2}\,\,
d\Omega_{7-p}^2\,\,,
\label{nueve}
\eeq
where $d\Omega_{7-p}^2$, which only depends on 
$\theta^1,\,\cdots\,,\theta^{7-p}$, is the line element of a $S^{7-p}$
sphere. We are going to consider static configurations of the
D(8-p)-brane which only depend on the polar angle $\theta$. Actually, we
shall restrict ourselves to the case in which the only non-vanishing
fields are $r(\theta)$, $x(\theta)$ and $A_0(\theta)$, $x$ being a
direction parallel to the background Dp-branes. It is straightforward
to verify that the action for such configurations can be written as:
\bear
S\,&=&\,T\,T_{8-p}\,\Omega_{7-p}\,\int d\theta\,\,
({\rm sin}\,\theta)^{7-p}\,\times\rc\rc
&&\times\,
\Bigl[\,-r^{7-p}\,H_p(r)\,
\sqrt{r^2\,+\,r\,'^{\,2}\,+\,{x\,'^{\,2}\over H_p(r)}\,
-\,F_{0,\theta}^2}\,+\,
(-1)^{p+1}\,(7-p)\,A_0\,R^{7-p}\,\Bigr]\,\,,\rc\rc
\label{diez}
\eear
where the prime denotes derivative with respect to $\theta$ and 
$T\,=\,\int\,dt$. In eq. (\ref{diez}) $\Omega_{7-p}$ is the volume of
the unit (7-p)-sphere, given by:
\beq
\Omega_{7-p}\,=\,{2\pi^{{8-p\over 2}}\over 
\Gamma\Bigl(\,{8-p\over 2}\Bigr)}\,\,.
\label{once}
\eeq
It is interesting at this point to remember that we are considering
static solutions of the gauge field that only depend on $\theta$. The
electric field for these configurations is:
\beq
E\,=\,F_{0,\theta}\,=\,-\partial_{\theta}\,A_0\,\,.
\label{doce}
\eeq
Let us now define the displacement field $D_p$ as:
\beq
D_p(\,\theta\,)\,\equiv\,{(-1)^{p+1}\over T\,T_{8-p}\,\Omega_{7-p}}\,
{\partial S\over \partial E}\,=\,(-1)^{p+1}
{({\rm sin}\,\theta)^{7-p}\,r^{7-p}\,H_p(r)\,E\over
\sqrt{r^2\,+\,r\,'^{\,2}\,+\,{x\,'^{\,2}\over H_p(r)}\,-\,E^2}}\,\,.
\label{trece}
\eeq
The extra factors appearing in the definition (\ref{trece}) have been
included for convenience. The Euler-Lagrange equation for $A_0$ implies
an equation for $D_p$ which is easy to find, namely:
\beq
{d\over d\theta}\,D_p(\,\theta\,)\,=\,
-(\,7\,-\,p\,)\,R^{7-p}\,(\,{\rm sin }\,\theta\,)^{7-p}\,\,.
\label{catorce}
\eeq
Notice that the right-hand side of eq. (\ref{catorce}) only depends
on $\theta$. Therefore (\ref{catorce}) can be integrated and, as a
result, one can obtain $D_p$ as a function of $\theta$. We shall work
out the explicit form of $D_p(\,\theta\,)$ at the end of this section.
At present we only need to use the fact that $D_p(\,\theta\,)$
satisfies eq. (\ref{catorce}). Actually, substituting 
$(\,7\,-\,p\,)\,R^{7-p}\,(\,{\rm sin }\,\theta\,)^{7-p}$ by 
$-\partial_{\theta}\,D_p$ in the Wess-Zumino term of the action and
integrating by parts, one can recast $S$ as:
\beq
S\,=\,-T\,U\,\,,
\label{quince}
\eeq
with $U$ given by:
\bear
U\,=\,T_{8-p}\,\Omega_{7-p}\,\int d\theta\,
&&\Biggl[\,\,(-1)^{p+1}\,E\,D_p(\theta)\,+\rc\rc
&&+\,({\rm sin}\,\theta)^{7-p}\,r^{7-p}\,H_p(r)\,
\sqrt{r^2\,+\,r\,'^{\,2}\,+\,{x\,'^{\,2}\over H_p(r)}\,-\,E^2}
\,\,\,\Biggr]\,\,.
\label{dseis}
\eear
Notice that, as is evident from their relation (\ref{quince}), $S$ and
$U$ give rise to the same Euler-Lagrange equation. Actually, since we
have eliminated $A_0$ in favor of $D_p$, we can regard $U$ as the
Legendre transform of $S$ and, therefore, $U$ can be considered as an
energy functional for the embedding of the D(8-p)-brane in the
Dp-brane background. The fields $E$ and $D_p$ are related, as is
obvious from eq. (\ref{trece}). It is not difficult to invert eq. 
(\ref{trece}) and get $E$ in terms of $D_p$. The result is:
\beq
E\,=\,(-1)^{p+1}\,
\sqrt{
{r^2\,+\,r\,'^{\,2}\,+\,{x\,'^{\,2}\over H_p(r)}\over
(D_p(\theta))^2\,+
[({\rm sin}\,\theta)^{7-p}\,r^{7-p}\,H_p(r)\,]^2}}
\,\,\,D_p(\theta)\,\,.
\label{dsiete}
\eeq
Using this relation we can eliminate $E$ from the expression of $U$:
\beq
U\,=\,T_{8-p}\,\Omega_{7-p}\,\int d\theta\,
\sqrt{r^2\,+\,r\,'^{\,2}\,+\,{x\,'^{\,2}\over H_p(r)}}\,\,\,
\sqrt{(D_p(\theta))^2\,+
[({\rm sin}\,\theta)^{7-p}\,r^{7-p}\,H_p(r)\,]^2}\,\,.
\label{docho}
\eeq
Recall that $D_p(\theta)$ is a known function of $\theta$ (see
below). Therefore eq. (\ref{docho}) gives the energy functional $U$ in
terms of $x(\theta)$ and $r(\theta)$. These functions must be
solutions of the Euler-Lagrange equations obtained from $U$. From the
study of the functional $U$ in several situations we will determine
the shape of the D(8-p)-brane embedding in the background geometry. Let
us consider, first of all, the near-horizon approximation, which is
equivalent to taking $a=0$ in the harmonic function (\ref{dos}). In
this case  $r^{7-p}\,H_p(r)\,=\,R^{7-p}$ and $U$ is given by:
\beq
U\,=\,T_{8-p}\,\Omega_{7-p}\,\int d\theta\,
\sqrt{r^2\,+\,r\,'^{\,2}\,+\,
{r^{7-p}\over R^{7-p}}\,\,x\,'^{\,2}}\,\,\,
\sqrt{(D_p(\theta))^2\,+({\rm sin}\,\theta)^{2(7-p)}\,R^{2(7-p)}\,}
\,\,.
\label{dnueve}
\eeq
By inspecting eq. (\ref{dnueve}) one easily concludes that $U$
transforms homogeneously under a simultaneous rescaling of the radial
and parallel coordinates of  the form:
\beq
(r,x)\,\rightarrow\,(\,\alpha\,r,\,\alpha^{{p-5\over 2}}\,x\,)\,\,,
\label{veinte}
\eeq
where $\alpha$ is a constant. It follows that the equations of motion
derived from $U$ are invariant under the transformation
(\ref{veinte}). Moreover, the homogeneous character of $U$ under the
transformation (\ref{veinte}) implies the following scaling law:
\beq
x\,\sim\,{R^{{7-p\over 2}}\over r^{{5-p\over 2}}}\,\,,
\label{vuno}
\eeq
where the power of $R$ has been determined by imposing dimensional
homogeneity of both sides of (\ref{vuno}). Eq.  (\ref{vuno}) is
precisely the holographic UV/IR relation found in ref.
\cite{holography}. According to eq. (\ref{vuno}), 
for $p<5$, large radial
distances  ($r\rightarrow\infty$) correspond to small values of the
parallel coordinate $x$. For $p=5$, the distance $x$ is insensitive to
changes in $r$, whereas, for $p=6$,  $x$ increases when $r$ grows. The
consequences of this relation for the correspondence between field
theories and near-horizon supergravities have been discussed in ref.
\cite{holography} (see also ref. \cite{holoothers}). 
In our approach we shall see that, indeed, the
$p=6$ case is special and the embedding of the D2-brane in the
D6-brane background geometry has new characteristics, which must be
studied separately.

Let us finish this section by giving the expressions of the
displacement fields $D_p$. The $\theta$-dependence of $D_p$ can be
obtained by integrating the right-hand side of eq. (\ref{catorce}).
The results one gets for the different cases are:

\bear
D_0(\,\theta\,)\,&=&\,
R^7\,\Bigl[\,{\rm cos}\,\theta\,(\,{\rm sin }^6\theta\,+\,
{6\over 5} \,{\rm sin }^4\,\theta\,+\,{8\over 5} \,
{\rm sin }^2\,\theta\,+\,{16\over 5}\,)\,+\,
{16\over 5}\,(\,2\nu\,-\,1\,)\,\Bigr]\,\,,
\rc\rc
D_1(\,\theta\,)\,&=&\,
R^6\,\Bigl[\,{\rm cos}\,\theta\,(\,{\rm sin }^5\theta\,+\,
{5\over 4} \,{\rm sin }^3\theta\,+\,{15\over 8} \,{\rm sin }\,\theta\,)
\,+\,{15\over 8} \,(\,\pi\nu\,-\,\theta\,)\,\Bigr]\,\,,
\rc\rc
D_2(\,\theta\,)\,&=&\,
R^5\,\Bigl[\,{\rm cos}\,\theta\,(\,{\rm sin }^4\theta\,
+\,{4\over 3} \,{\rm sin }^2\theta\,+\,
{8\over 3}\,)\,+\,{8\over 3}\,(\,2\nu\,-\,1\,)\Bigr]\,\,,
\rc\rc
D_ 3(\,\theta\,)\,&=&\,
R^4\,\Bigl[\,{\rm cos}\,\theta\,(\,{\rm sin }^3\theta\,
+\,{3\over 2} \,{\rm sin }\,\theta\,)\,+\,
{3\over 2} \,(\,\pi\nu\,-\,\theta\,)\,\Bigr]\,\,,\label{vdos}\\\rc
D_ 4(\,\theta\,)\,&=&\,R^3\,\Bigl[\,{\rm cos}\,\theta\,
(\,{\rm sin}^2\theta\,+\,2\,)\,+\,2\,(\,2\nu\,-\,1\,)\Bigr]\,\,,
\rc\rc
D_ 5(\,\theta\,)\,&=&\,R^2\,\Bigl[\,{\rm cos}\,\theta\,
{\rm sin }\,\theta\,+\,\pi\nu\,-\,\theta\,\Bigr]\,\,,
\rc\rc
D_ 6(\,\theta\,)\,&=&\,R\,\Bigl[\,{\rm cos}\,\theta\,
+\,2\nu\,-\,1\,\Bigr]\,\,,
\nonumber
\eear
where we have parametrized the additive constant of integration by
means of a parameter $\nu$. We have chosen the $\theta$-independent
term in $D_p(\theta)$ in such a way that the value of the displacement
field at $\theta=\pi$ is given by:
\beq
D_p(\pi)\,=\,-2\sqrt\pi\,{\Gamma\Bigl(\,{8-p\over 2}\Bigr)\over
\Gamma\Bigl(\,{7-p\over 2}\Bigr)}\,\,R^{7-p}\, (\,1\,-\,\nu\,)\,\,.
\label{vtres}
\eeq
Eq. (\ref{vtres}) allows to give a precise meaning to the parameter
$\nu$ \cite{CGS1}. 
In fact, as we shall verify below, there exist solutions for
which $r\rightarrow\infty$ and $x'\rightarrow 0$ when
$\theta\rightarrow\pi$ in a way that simulates a `flux tube' attached
to the D(8-p)-brane. The `tension' (\ie\ the energy per unit radial
distance) for one of these spikes can be obtained from the expression
of $U$ in eq. (\ref{docho}). Actually, as $\theta\rightarrow\pi$ one
can check that the terms with $r\,'$ dominate over the other terms in
the first square root in eq. (\ref{docho}), whereas the second square
root can be approximated by $|\,D_p(\pi)\,|$. 
As a result \cite{CGS1}, the
tension of the spike is $T_{8-p}\,\Omega_{7-p}\,|\,D_p(\pi)\,|$. Using
the value of $D_p(\pi)$ given in eq. (\ref{vtres}) and the fact 
that (see eqs. (\ref{tres}), (\ref{siete}) and (\ref{once})):
\beq
T_{8-p}\,\Omega_{7-p}\,R^{7-p}\,=\,
{NT_f\over 2\sqrt{\pi}}\,
{\Gamma\Bigl(\,{7-p\over 2}\Bigr)\over
\Gamma\Bigl(\,{8-p\over 2}\Bigr)}\,\,,
\label{vcuatro}
\eeq
we conclude
that the tension of the $\theta=\pi$ spike is  $(1-\nu)\,N\,T_f$,
where $T_f$ is the tension of the fundamental string written in eq. 
(\ref{cuatro}). This result implies that we can interpret the 
$\theta=\pi$ tube as a bundle of $(1-\nu)\,N$ fundamental strings
which, from the gauge theory point of view,  corresponds to a baryon
formed by $(1-\nu)\,N$ quarks. It is clear that, although in the
classical theory $\nu$ is a continuous parameter, upon quantization
$\nu$ should be a multiple of $1/N$ taking values in the range 
$0\le\nu\le1$. 

Let us finally point out that there exist other solutions for which 
$r\rightarrow\infty$ and $x'\rightarrow 0$ when
$\theta\rightarrow 0$. The asymptotic tension for these solutions is 
$T_{8-p}\,\Omega_{7-p}\,|\,D_p(0)\,|$. The value of $D_p(\theta)$ at
$\theta=0$ can be obtained from the values given in eq. (\ref{vdos}).
The result is:
\beq
D_p(0 )\,=\,2\sqrt\pi\,{\Gamma\Bigl(\,{8-p\over 2}\Bigr)\over
\Gamma\Bigl(\,{7-p\over 2}\Bigr)}\,\,R^{7-p}\, \nu\,\,.
\label{vcinco}
\eeq
Using again eq. (\ref{vcuatro}), one can verify that 
$T_{8-p}\,\Omega_{7-p}\,|\,D_p(0)\,|\,=\,\nu\,N\,T_f$, which
corresponds to a bundle of $\nu\,N$ fundamental strings. This fact
provides an interpretation of $\nu$ for this second class of
solutions.

\setcounter{equation}{0}
\section{BPS conditions}
\medskip
In the remaining of this paper we are going to study solutions of the
equations of motion of the brane probes which are not extended in the
directions parallel to the background branes. This amounts to take
$x\,'=0$ (\ie\ $x(\theta)\,=\,$ constant) in our previous equations.
From the expression (\ref{docho}) of $U$ we can get the differential
equation which determines $r$ as a function of $\theta$. Indeed, the
Euler-Lagrange equation derived from $U$ is:

\bear
&&{d\over d\theta}\,\Biggr[\,{r'\over \sqrt{r^2\,+\,r\,'^{\,2}}}
\sqrt{(D_p(\theta))^2\,+[({\rm sin}\,\theta)^{7-p}
\,r^{7-p}\,H_p(r)\,]^2}
\,\Biggr]\,=\,\rc\rc\rc
&&=\,{r\over \sqrt{r^2\,+\,r\,'^{\,2}}}\,\,
\sqrt{(D_p(\theta))^2\,+[({\rm sin}\,\theta)^{7-p}\,
r^{7-p}\,H_p(r)\,]^2}
\,+\,\label{vseis}\\\rc\rc
&&+\,(7-p)a\,\,{\sqrt{r^2\,+\,r\,'^{\,2}}\over r}\,\,
{(\,r\,{\rm sin }\,\theta\,)^{2(7-p)}\,H_p(r)\over
\sqrt{(D_p(\theta))^2\,+[({\rm
sin}\,\theta)^{7-p}\,r^{7-p}\,H_p(r)\,]^2}}\,\,.
\rc\rc
\nonumber
\eear
Trying to obtain a solution for such a complicate second-order
differential equation seems, a priori, hopeless. A possible strategy
to solve eq. (\ref{vseis}) consists of finding a first integral for
this system. Notice that the integrand of $U$ depends explicitly on
$\theta$ (see eq.  (\ref{docho})) and, therefore,  there is no first
integral associated to the invariance under shifts of $\theta$ by a
constant. Nevertheless, we will be able to find a first-order equation
such that any function $r(\theta)$ satisfying it is a solution of the
equation  (\ref{vseis}). This first-order condition is much simpler
than the Euler-Lagrange equation (\ref{vseis}) and, indeed, we will be
able to solve it analytically, both in the near-horizon ($a=0$) and
asymptotically flat ($a=1$) geometries. For $p=3$ and $a=0$, the first
order equation was found by Imamura 
\cite{Imamura} as a BPS \cite{BPS} condition for the
implementation of supersymmetry in the worldvolume theory of a
D5-brane propagating in a D3 background. This condition was
subsequently extended to ($p=3$, $a=1$)  and ($p=4$, $a=0$) in
refs. \cite{CGS1} and \cite{CGS2} respectively. 
In order to generalize these results,
 we shall  follow here the approach 
of ref. \cite{Craps} (see also ref. \cite{Gomis}),
where the BPS equation was found by requiring the saturation of a
certain bound on the energy functional. 

Following ref. \cite{Craps}, let us define the quantity:
\beq
\Delta_p(\,r\,,\,\theta\,)\,\equiv\,
r^{7-p}\,H_p(r)\,\,(\,{\rm sin }\,\theta\,)^{7-p}\,\,.
\label{vsiete}
\eeq
In terms of $\Delta_p(\,r\,,\,\theta\,)$, we define the function 
$f_p(r, \theta)$ as follows:
\beq
f_p(r, \theta)\,\equiv\,
{\Delta_p(\,r\,,\,\theta\,)\,{\rm sin }\,\theta\,\,+\,
D_p(\,\theta\,)\,{\rm cos }\,\theta\over
\Delta_p(\,r\,,\,\theta\,)\,{\rm cos }\,\theta
\,-\,D_p(\,\theta\,)\,{\rm sin }\,\theta}\,\,.
\label{vocho}
\eeq
Making use of the function $f_p$ it is easy to verify that the energy
can be put in terms of a square root of a sum of squares. Indeed, it
is a simple exercise to demonstrate that $U$ can be written as:
\bear
U\,&=&\,
T_{8-p}\,\Omega_{7-p}\,\int d\theta\,
\,\,\times\rc\rc
&&\times\,\sqrt{{\cal Z}^2\,+\,r^2\,
\Bigl[\,\Delta_p(\,r\,,\,\theta\,)\,
{\rm cos }\,\theta-\,
D_p(\theta)\,{\rm sin }\,\theta\,\Bigr]^2\,\Bigl[\,
{r'\over r}-f_p(r, \theta)\,\Bigr]^2\,\,,
}
\label{vnueve}
\eear
where ${\cal Z}$ is given by:
\beq
{\cal Z}\,=\,r\,
\Bigl[\,\Delta_p(\,r\,,\,\theta\,)\,
{\rm cos }\,\theta-\,
D_p(\theta)\,{\rm sin }\,\theta\,\Bigr]\,
\Bigl[\,1\,+\,{r'\over r}\,f_p(r, \theta)\,\Bigr]\,\,.
\label{treinta}
\eeq
In view of eq. (\ref{vnueve}), it is clear that the energy of the
D(8-p)-brane is bounded as:
\beq
U\,\ge\,T_{8-p}\,\Omega_{7-p}\,\int d\theta\,
\Bigl|\,{\cal Z}\,\Bigr|\,\,.
\label{tuno}
\eeq
This bound is saturated when $r(\theta)$ satisfies the following
first-order differential equation:
\beq
{r'\over r}\,=\,
f_p(r, \theta)\,\,.
\label{tdos}
\eeq
It is straightforward to prove that any function $r(\theta)$ that
satisfies eq. (\ref{tdos}) is also a solution of the Euler-Lagrange
equation (\ref{vseis}). Notice that the condition (\ref{tdos})
involves the displacement field $D_p(\theta)$ (see eq. (\ref{vocho})).
The expressions of the $D_p$'s have been given at the end of section 2
(eq. (\ref{vdos})). However, in order to demonstrate that the
solutions of eq. (\ref{tdos}) verify the equation of motion
(\ref{vseis}), we do not need to use the explicit form of 
$D_p(\theta)$. The only property required for this proof is the value
of the derivative of $D_p(\theta)$, given in eq. (\ref{catorce}).
Moreover, as pointed out in ref. \cite{Craps} for 
the $p=3$ case, ${\cal Z}$ can be written as a total derivative:

\beq
{\cal Z}\,=\,
{d\over d\theta}\,\,
\Bigl[\,D_p(\theta)\,r\,{\rm cos }\,\theta\,+\,
\Bigl(\,{a\over 8-p}\,+\,{R^{7-p}\over r^{7-p}}\,\Bigr)
\,\,(\,r\,{\rm sin }\,\theta\,)^{8-p}\,\Bigr]\,\,.
\label{tseis}
\eeq
It is important to stress the fact that eq. (\ref{tseis}) can be
proved without using the condition (\ref{tdos}) or the
Euler-Lagrange equation (\ref{vseis}) (only eq. (\ref{catorce}) has
to be used). Eq. (\ref{tseis}) implies that the bound (\ref{tuno})
does not depend on the detailed form of the function $r(\theta)$.
Actually, as a consequence of eq. (\ref{tseis}), only the boundary
values of $r(\theta)$ matter when one evaluates the right-hand side of
eq. (\ref{tuno}). The functions $r(\theta)$ which solve eq. (\ref{tdos})
correspond to those  D(8-p)-brane embeddings which, for given
boundary conditions, have minimal energy. Due to this fact we shall
refer to eq. (\ref{tdos}) as the BPS condition. 
The search for the solutions of the BPS equation  and their
interpretation will be the subject of the next three sections.

\setcounter{equation}{0}
\section{The near-horizon solution for $\bf {p\le 5}$}
\medskip

In this section we are going to solve the BPS first-order differential
equation in the near-horizon geometry for $p\le 5$. It will become
clear in the process of finding the solution that the $p=6$ case is
singled out (as expected from the holographic relation (\ref{vuno})).
This $p=6$ case will be discussed separately in section 5.

Let us start by defining the function $\Lambda_p(\,\theta\,)$ by means
of the equation:

\beq
\Lambda_p(\,\theta\,)\,\equiv\,{1\over R^{7-p}}\,
\Bigl[\,R^{7-p}\,(\,{\rm sin }\,\theta\,)^{6-p}\,{\rm cos }\,\theta
\,-\,D_p(\,\theta\,)\,\Bigr]\,\,.
\label{tnueve}
\eeq
The right-hand side of eq. (\ref{tnueve}) is known and, thus, 
$\Lambda_p(\,\theta\,)$ is a known function of $\theta$ whose explicit
expression can be obtained by substituting the values of
$D_p(\,\theta\,)$, given in eq. (\ref{vdos}), in 
 (\ref{tnueve}). Moreover, the BPS condition (\ref{tdos}) for $a=0$
can be written in terms of $\Lambda_p(\,\theta\,)$ as:

\beq
{r'\over r}\,=\,{({\rm sin }\,\theta\,)^{6-p}\,-\,
\Lambda_p(\,\theta\,)\,{\rm cos }\,\theta\over
\Lambda_p(\,\theta\,)\,{\rm sin }\,\theta}\,\,.
\label{cuarenta}
\eeq
A key point in what follows is that $\Lambda_p(\,\theta\,)$ has a
simple derivative, which can be easily obtained from eq.
(\ref{catorce}), namely:

\beq
{d\over d\theta}\,\Lambda_p(\,\theta\,)\,=\,
(\,6\,-\,p)\,(\,{\rm sin }\,\theta\,)^{5-p}\,\,.
\label{cuno}
\eeq
By inspecting  the right-hand side of this equation we observe that
$p=6$ is a special case. Indeed, when $p\not= 6$,  we can represent the 
$({\rm sin }\,\theta\,)^{6-p}$ term appearing in the BPS condition as:

\beq
({\rm sin }\,\theta\,)^{6-p}\,=\,{{\rm sin }\,\theta\over 6-p}\,
\,\,{d\over d\theta}\,\,\Lambda_p(\,\theta\,)\,\,,
\,\,\,\,\,\,\,\,\,\,\,\,\,\,\,\,\,\,\,\,\,\,\,\,\,\,\,\,
(p\not= 6)\,\,.
\label{cdos}
\eeq
After doing this, the right-hand side of eq. (\ref{cuarenta}) is
immediately recognized as a logarithmic derivative and, as a
consequence, the BPS condition can be readily integrated. The result
one arrives at is:

\beq
r\,(\,\theta\,)\,=\,C\,\,
{\,\Bigl[\,\Lambda_p(\,\theta\,)\,\Bigr]^{{1\over 6-p}}\over
{\rm sin }\,\theta}\,\,,
\label{ctres}
\eeq
where $C$ is a positive constant. For $p<5$ there is a fractional power
of $\Lambda_p(\,\theta\,)$ in the right-hand side of eq.
(\ref{ctres}) and, thus, the solution we have found only makes sense
for those values of $\theta$ such that
$\Lambda_p(\,\theta\,)\,\ge 0$. Moreover, it is clear from eq. 
(\ref{cuno}) that:

\beq
{d\Lambda_p\over d\theta}\,\ge\,0\,\,,
\,\,\,\,\,\,\,\,\,\,\,\,\,\,\,\,\,\,\,\,\,\,\,\,\,\,\,\,
(\,p<6\,,\,\,\,\,0\,\le\,\theta\,\le \,\pi\,)\,\,,
\label{ccuatro}
\eeq
and, therefore, $\Lambda_p(\,\theta\,)$ is a monotonically increasing
function in the interval $0\le\theta\le\pi$. The values of 
$\Lambda_p(\,\theta\,)$ at $\theta\,=\,0,\pi$ are:

\bear
\Lambda_p(0)\,&=&-{D_p(0)\over R^{7-p}}\,=\,-\,
2\sqrt\pi\,{\Gamma\Bigl(\,{8-p\over 2}\Bigr)\over
\Gamma\Bigl(\,{7-p\over 2}\Bigr)}\,\, \nu\,\le\,0\,\,,
\rc\rc
\Lambda_p(\pi)\,&=&-{D_p(\pi)\over R^{7-p}}\,=\,
2\sqrt\pi\,{\Gamma\Bigl(\,{8-p\over 2}\Bigr)\over
\Gamma\Bigl(\,{7-p\over 2}\Bigr)}\,\,
(\,1\,- \nu\,)\,\ge\,0\,\,,
\label{ccinco}
\eear
where, in order to establish the last inequalities, we have used the
fact that $0\le\nu\le 1$. From this discussion it follows that there
exists a unique value $\theta_0$ of the polar angle such that:
\begin{figure}
\centerline{\hskip -.8in \epsffile{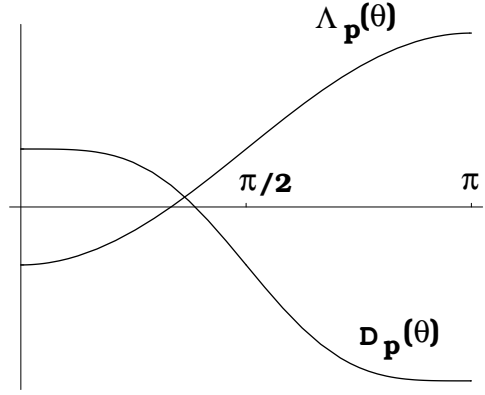}}
\caption{The functions $\Lambda_p(\theta)$ and $D_p(\theta)$ for
$0\le\theta \le\pi$. For illustrative purposes we have plotted these
functions for $p=4$ and $\nu=1/4$.}
\label{fig1}
\end{figure}

\beq
\Lambda_p\,(\,\theta_{0}\,)\,=\,0\,\,.
\label{cseis}
\eeq
The form of the function $\Lambda_p\,(\,\theta\,)$ has been
displayed in figure 1. It is clear that the solution (\ref{ctres}) is
valid for 
$\theta_0\le\theta\le\pi$. Notice that $\theta_0$ depends on $\nu$ and,
actually, is a monotonically increasing function of $\nu$. In fact,
from eq. (\ref{ccinco}) one concludes that $\theta_0\,=\,0$ for 
$\nu\,=\,0$, whereas $\theta_0\,=\,\pi$ for $\nu\,=\,1$.

For $p=5$ the function  appearing in the right-hand side of eq.  
(\ref{ctres}) is:
\beq
\Lambda_5(\,\theta\,)\,=\,\theta\,-\,\pi\,\nu\,\,.
\label{csiete}
\eeq
Thus, our solution in this case  is:
\beq
r\,(\,\theta\,)\,=\,C\,\,
{\,\theta\,-\,\pi\,\nu\over
{\rm sin }\,\theta}\,\,,
\,\,\,\,\,\,\,\,\,\,\,\,\,\,\,\,\,\,\,\,\,\,\,\,\,\,\,\,
(\,p\,=\,5\,)\,\,.
\label{cocho}
\eeq
As the radial coordinate $r$ must be non-negative, from eq.
(\ref{cocho}) one immediately concludes that, also in this 
$p=5$ case, $\theta_0\le\theta\le\pi$, where now $\theta_0$ depends
linearly on $\nu$, namely:
\beq
\theta_{0}\,=\,\pi\,\nu\,\,,
\,\,\,\,\,\,\,\,\,\,\,\,\,\,\,\,\,\,\,\,\,\,\,\,\,\,\,\,
(\,p\,=\,5\,)\,\,.
\label{cnueve}
\eeq
The solution we have found coincides with 
the one obtained in refs. \cite{CGS1, CGS2} 
for $p=3,4$. Our result (\ref{ctres}) generalizes these solutions for
any $p<6$ (the solution for $p=6$ will be given in the next section).
By inspecting eq. (\ref{ctres}) it is easy to conclude that, for
$\nu\not=1$, $r(\theta)$ diverges when $\theta\rightarrow\pi$.
Actually, when $\theta\approx\pi$ and $\nu\not=1$, $r(\theta)$ behaves
as:
\beq
r(\,\theta\,)\,\approx\,
{C\,\Bigl[\,\Lambda_p(\,\pi\,)\,\Bigr]^{{1\over 6-p}}\over
\pi\,-\,\theta}\,\,,
\,\,\,\,\,\,\,\,\,\,\,\,\,\,\,\,\,\,\,\,\,\,\,\,\,\,\,\,
(\,\nu\,\not=\,1\,)\,\,,
\label{cincuenta}
\eeq
which corresponds to a ``tube" of radius 
$C\,\Bigl[\,\Lambda_p(\,\pi\,)\,\Bigr]^{{1\over 6-p}}$. We will check
below that the energy of these tubes corresponds to a bundle of 
$(1-\nu)N$ fundamental strings emanating from the D(8-p)-brane. When 
$\nu\not=0$ the critical angle $\theta_0$ is greater than zero and, as
a consequence, eq. (\ref{ctres}) gives $r(\theta_0)\,=\,0$. Actually,
when $\nu\not=0$, the solution has the shape represented in figure 2a.
On the contrary, when $\nu=0$, the function $r(\theta)$ takes a
non-vanishing value at the critical value $\theta_0\,=\,0$. Notice
that, in this last case, both numerator and denominator on the
right-hand side of eq. (\ref{ctres}) vanish. The behaviour of 
$\Lambda_p(\,\theta\,)$ near $\theta=0$ can be obtained by a Taylor
expansion. The successive derivatives of $\Lambda_p(\,\theta\,)$ can be
easily computed from the value of its first derivative, given in eq. 
(\ref{cuno}).  It is easy to prove that the first non-vanishing
derivative of $\Lambda_p(\,\theta\,)$ at $\theta=0$ is:

\begin{figure}
\centerline{\hskip -.8in \epsffile{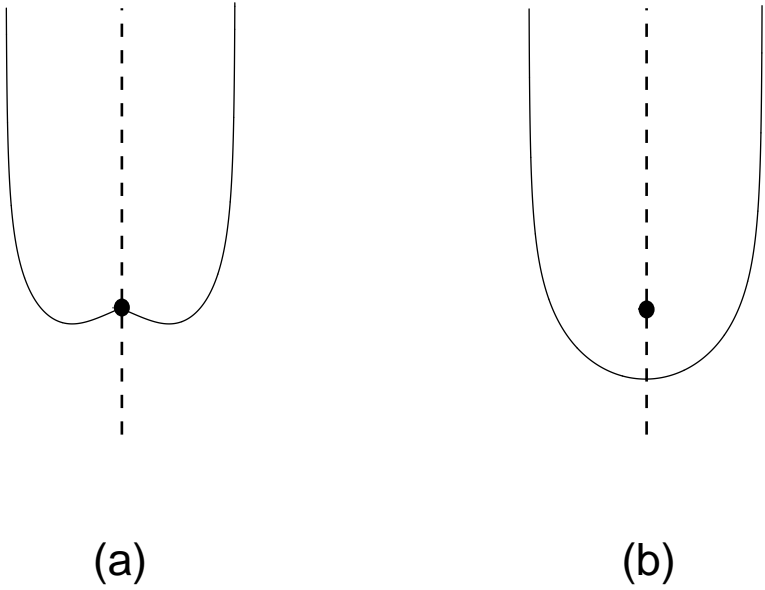}}
\caption{Plot of the near-horizon solution (\ref{ctres}) for
$(p=3,\nu=1/4)$ (a) and $(p=3,\nu=0)$ (b). The discontinuous line
represents the polar axis ($\theta=\pi$ at the top). The dot
corresponds to the origin $r=0$. The background branes are located at
the origin and extend in the directions orthogonal to the plane of the
figure. }
\label{fig2}
\end{figure}

\beq
{d^{\,6-p}\over d\theta^{\,6-p}}\,\Lambda_p(\,\theta\,)\,\,
\Bigg|_{\theta=0}\,=\,(\,6\,-\,p\,)!\,\,.
\label{cinuno}
\eeq
Moreover, the $(7-p)^{{\rm th}}$ derivative of 
$\Lambda_p(\,\theta\,)$ at $\theta=0$ vanishes, \ie:

\beq
{d^{\,7-p}\over d\theta^{\,7-p}}\,\Lambda_p(\,\theta\,)\,\,
\Bigg|_{\theta=0}\,=\,0\,\,.
\label{cidos}
\eeq
Therefore, the expression of $\Lambda_p(\,\theta\,)$ for values of 
$\theta$ close to zero takes the form:

\beq
\Lambda_p(\,\theta\,)\,\approx\,\Lambda_p(\,0\,)\,+\,
\theta^{\,6-p}\,+\,c_p\,\theta^{\,8-p}\,+\,
\cdots\,\,,
\label{citres}
\eeq
where $c_p$ is a non-vanishing coefficient. Substituting this
expansion in the right-hand side of eq. (\ref{ctres}) for $\nu=0$
leads to the conclusion that:
\beq
r(\,0\,)\,=\,C\,,
\,\,\,\,\,\,\,\,\,\,\,\,\,\,\,\,\,\,\,\,\,\,\,\,\,\,\,\,
(\,\nu\,=\,0\,)\,\,,
\label{cicuatro}
\eeq
\ie\ $r(0)\not=0$ for $\nu=0$, as claimed above. Moreover, $r'(\,0\,)$
vanishes in this case:
\beq
r'(\,0\,)\,=\,0\,,
\,\,\,\,\,\,\,\,\,\,\,\,\,\,\,\,\,\,\,\,\,\,\,\,\,\,\,\,
(\,\nu\,=\,0\,)\,\,.
\label{cicinco}
\eeq
The shape of the solution for $\nu=0$ has been plotted in figure 2b.
Notice that the $\nu=0$ tube corresponds to the ordinary baryon with
$N$ quarks.

The solution of the near-horizon BPS condition written in  eq.
(\ref{ctres}) is not the only one. Following 
ref. \cite{CGS1}, we can construct
a new solution, valid for values of $\theta$ in the range 
$0\le\theta\le\theta_0$, in which $\Lambda_p(\,\theta\,)\le 0$, as
follows:

\beq
\tilde r\,(\,\theta\,)\,=\, C\,\,
{\,\Bigl[\,-\Lambda_p(\,\theta\,)\,\Bigr]^{{1\over 6-p}}\over
{\rm sin }\,\theta}\,\,.
\label{ciseis}
\eeq
Eq. (\ref{ciseis}) describes a spike at $\theta\,=\,0$. We shall refer
to the solution (\ref{ciseis}) as a ``lower tube" solution, in
contrast to the ``upper tube" solution of eq. (\ref{ctres}). Actually,
these two types of solutions are related. In order to verify this
fact, let us point out that the functions $D_p(\,\theta\,)$ and 
$\Lambda_p(\,\theta\,)$  change their sign under the transformation 
$\theta\rightarrow\pi-\theta$, $\nu\rightarrow1-\nu$:

\bear
D_p(\,\theta\,;\,\nu\,)\,&=&\,-
D_p(\,\pi\,-\theta\,;\,1\,-\nu\,)\,\,,\rc\rc
\Lambda_p(\,\theta\,;\,\nu\,)\,&=&\,-
\Lambda_p(\,\pi\,-\theta\,;\,1\,-\nu\,)\,\,,
\label{cisiete}
\eear
and, therefore, the solutions (\ref{ctres}) and (\ref{ciseis}) are
simply related, namely:

\beq
\tilde r\,(\,\theta\,;\,\nu\,)\,=\,
r\,(\,\pi\,-\,\theta\,;\,1\,-\,\nu\,)\,\,.
\label{ciocho}
\eeq
It follows from this relation that the lower flux tubes correspond to 
$\nu N$ quarks. 

Let us now evaluate, following ref. \cite{Craps},  the energy for the
two types of near-horizon solutions we have found. These solutions
saturate the bound (\ref{tuno}) and, thus, the energy is precisely the
right-hand side of this equation. Moreover, when $r$ is a solution of
eq. (\ref{tdos}), ${\cal Z }$ can be written as:

\beq
{\cal Z }\,=\,r\,[\,\Delta_p(\,r\,,\,\theta\,)\,\,
{\rm cos }\,\theta\,-\,D_p(\,\theta\,)\,
{\rm sin }\,\theta\,]\,\Bigl[\,1\,+\,
\Bigl(\,{r\,'\over r}
\Bigr)^2\,\,\Bigr]\,\,.
\label{cinueve}
\eeq
It is obvious from this equation that, for our
solutions,  the sign of ${\cal Z }$  is just the sign of 
$\Delta_p(\,r\,,\,\theta\,)\,\,{\rm cos }\,\theta\,-\,D_p(\,\theta\,)\,
{\rm sin }\,\theta$. Moreover, using the definitions of $\Delta_p$ 
({eq.~(\ref{vsiete})) and $\Lambda_p$ (eq.~(\ref{tnueve})), one
can prove that: 
\beq
\Delta_p(\,r\,,\,\theta\,)\,\,
{\rm cos }\,\theta\,-\,D_p(\,\theta\,)\,
{\rm sin }\,\theta\,=\,a\,r^{7-p}\,
(\,{\rm sin }\,\theta\,)^{7-p}\,{\rm cos }\,\theta\,+\,
R^{7-p}\,\Lambda_p(\,\theta\,)\,{\rm sin }\,\theta\,\,,
\label{sesenta}
\eeq
where we have used the general harmonic function 
(\ref{dos}). In the near-horizon case ($a=0$) the first term on the
right-hand side of eq. (\ref{sesenta}) is absent and, therefore, the
sign of $\Delta_p(\,r\,,\,\theta\,)\,\,{\rm cos }\,
\theta\,-\,D_p(\,\theta\,)\,
{\rm sin }\,\theta$ is just the sign of $\Lambda_p(\theta)$. In the
solutions (\ref{ctres}) and (\ref{ciseis}), the angle $\theta$ can
take values in a range such that $\Lambda_p(\,\theta\,)$ has a well
defined sign. Therefore, we can write:

\beq
\int d\theta\,
\Bigl|\,{\cal Z}\,\Bigr|\,=\,
\,\,\Bigl|\,\int d\theta\,{\cal Z}\,\Bigr|\,\,.
\label{tcuatro}
\eeq
If we now define $Z$ as:
\beq
Z\,\equiv\,T_{8-p}\,\Omega_{7-p}\,\int d\theta\,{\cal Z}\,\,,
\label{ttres}
\eeq
it is immediate that the energy of our near-horizon BPS solutions is:

\beq
U_{BPS}\,=\,|\,Z\,|\,\,.
\label{tcinco}
\eeq

In order to evaluate $Z$ for the near-horizon solutions, we make use
of the representation (\ref{tseis}) of ${\cal Z}$ as a total
derivative. Recall that $\theta$ varies in the range 
$\theta_i\le\theta\le\theta_f$ where $\theta_i=\theta_0$ 
($\theta_i=0$) and $\theta_f=\pi$ 
($\theta_f=\theta_0$) for the upper (lower) tube solution. It follows
that $Z$ can be written as the sum
\footnote{The labels $s$  and $gs$ refer to ``spike" and  ``ground
state", according to the interpretation given to these two
contributions in ref. \cite{Craps}.}: 

\beq
Z\,=\,Z_s\,+\,Z_{gs}\,\,,
\label{tsiete}
\eeq
where $Z_s$ and $Z_{gs}$ are given by:

\bear
Z_s\,&=&T_{8-p}\,\Omega_{7-p}\,D_p(\theta)\,
r(\theta)\,{\rm cos }\,\theta\,\Bigr|_{\theta_i}^{\theta_f}\,\,,\rc\rc
Z_{gs}\,&=&T_{8-p}\,\Omega_{7-p}\,
R^{7-p}\,\,
r(\theta)\, \,\,(\,{\rm sin }\,\theta\,)^{8-p}\,
\Bigr|_{\theta_i}^{\theta_f}\,\,.
\label{tocho}
\eear
It is clear that $Z$ only depends on the values of $r(\theta)$ at the
boundaries $\theta=\theta_i$ and  $\theta=\theta_f$. In this sense we
can regard $Z$ as a topological quantity which, for fixed boundary
conditions, is invariant under local variations of the fields
\cite{Craps}.

Let us now compute $Z_{gs}$. After substituting the solutions
(\ref{ctres}) and (\ref{ciseis}) of the BPS equations, we get: 

\beq
Z_{gs}\,=\,\,T_{8-p}\,\Omega_{7-p}\,R^{7-p}\,C\,
(\,{\rm sin }\,\theta)^{7-p}\,
\Bigl[\,\pm\Lambda_p(\,\theta\,)\,\Bigr]^{{1\over 6-p}}
\,\,\Biggr|_{\theta_i}^{\theta_f}\,\,,
\label{suno}
\eeq
where the $+$ ($-$ ) sign corresponds to the upper (lower) tube
solution. The angles $\theta_i$ and $\theta_f$ can take the values
$0$, $\pi$ and $\theta_0$. For $\theta=0,\pi$ the right-hand side of
eq. (\ref{suno}) vanishes due to the $(\,{\rm sin }\,\theta)^{7-p}$
factor, whereas $\theta=\theta_0$ gives also a vanishing contribution
because $\Lambda_p(\,\theta_0\,)=0$ (see eq. (\ref{cseis})). In
conclusion, we can write:
\beq
Z_{gs}\,=\,0\,\,.
\label{sdos}
\eeq
In the same way one can obtain the values of $Z_s$. One can check in
this case that the contribution of $\theta=\theta_0$ to the right-hand
side of the first equation in (\ref{tocho}) vanishes and, as a
consequence, only $\theta=0,\pi$ contribute. After using eqs. 
(\ref{vtres})-(\ref{vcinco}), one gets the result:

\beq
Z_s\,=\,
\cases{(1\,-\,\nu\,)\,N\,T_f\,L\,,&\,\,\,\,\,(upper tube)\,\,,\cr\cr
        -\,\nu\,N\,T_f\,L\,,&\,\,\,\,\,(lower tube)\,\,,\cr}
\label{stres}
\eeq
where $L=r(\pi)$ ($L=r(0)$) for the upper (lower) tube solution. It is
now evident that $U_{BPS}\,=\,|\,Z_s\,|$ is equal to the energy of a
bundle of $(1-\nu)N$ or $\nu N$ fundamental strings. Notice that this
result is the same as the one found at the end of section 2 for the
energy of the $\theta=0$ and $\theta=\pi$ spikes. Let us finally
mention that in ref. \cite{Craps} it has been argued in favor of
interpreting
$Z_s$ as a central charge in the worldvolume supersymmetry algebra.

\setcounter{equation}{0}
\section{The near-horizon D6-D2  system}
\medskip

In this section we are going to integrate the BPS condition for $p=6$.
Notice that, according to  the definition (\ref{tnueve}) and the
expression of $D_6(\theta)$ given in eq. (\ref{vdos}),
$\Lambda_6\,(\,\theta\,)$ is constant, \ie:

\beq
\Lambda_6\,(\,\theta\,)\,=\,1\,-\,2\,\nu\,\,.
\label{scuatro}
\eeq
For $p=6$ we cannot use eq. (\ref{cdos}) and, hence, we have to deal
directly with eq. (\ref{cuarenta}). Actually, eq. (\ref{cuarenta})
makes sense only  for $\nu\not=1/2$ since
$\Lambda_6\,(\,\theta\,)$, and therefore the denominator of eq.
(\ref{cuarenta}) vanishes identically for $\nu={1\over 2}$. 
With this restriction\footnote{ The fact that
the $p=6$, $\nu=1/2$ BPS condition  is ill-defined is an artifact of
the near-horizon approximation.}, the $p=6$ BPS condition is:

\beq
{r'\over r}\,=\,{1\over 1\,-\,2\nu}\,\,
{1\over {\rm sin }\,\theta}\,-\,
{{\rm cos }\,\theta\over {\rm sin }\,\theta}\,\,.
\label{scinco}
\eeq
It is not difficult to realize that the right-hand side of eq. 
(\ref{scinco}) can be written as a total derivative, namely:

\beq
{r'\over r}\,=\,{d\over d\theta}\,\,
\Biggl[\,\,{\rm log}\,\,\Bigl[\,\,
{\Bigl(\,{\rm tan}\,{\theta\over 2}\,\,\Bigr)^{{1\over 1-2\nu}}
\over {\rm sin }\,\theta}\,\,\Bigr]\,\,\Biggr]\,\,.
\label{sseis}
\eeq
Therefore, the BPS equation can be immediately integrated. It is
convenient to put the result in the form:

\beq
r(\,\theta\,)\,=\,A\,\,\,
{\,\,\Bigl[\,
{\rm sin }\,{\theta\over 2}\Bigr]^{{2\nu\over 1\,-\,2\nu}}\over 
\,\,\Bigl[\,
{\rm cos }\,{\theta\over 2}\Bigr]^{{2(\,1-\nu\,)
\over 1\,-\,2\nu}}
}\,\,,
\label{ssiete}
\eeq
where $A$ is a positive constant. The nature of the solution 
(\ref{ssiete}) depends critically on the value of $\nu$.  First of
all, the range of values of $\theta$ is not restricted, \ie\ 
$0\le\theta\le\pi$. If $0\le\nu<1/2$, the 
${\rm cos }\,(\theta/2)$ term present in the denominator of 
(\ref{ssiete}) makes $r\rightarrow\infty$  at $\theta\approx\pi$,
whereas for  $1/2<\nu\le 1$ the solution diverges at 
$\theta\approx 0$. Actually, the solution for 
$0\le\nu<1/2$ is related to the one for $1/2<\nu\le 1$ by means of the
equation:

\beq
r(\,\theta\,;\,\nu\,)\,=\,
r(\,\pi\,-\,\theta\,;\,1-\nu\,)\,\,.
\label{socho}
\eeq
Eq. (\ref{socho}) can be checked easily from the expression
(\ref{ssiete}). Notice that for 
$0<\nu<1/2$ ($1/2<\nu<1$) the D2 brane  passes through the point
$r=0$. This behaviour is similar to the $p<6$ case. However, the
present solution differs substantially  from those studied in section
4. In order to make  this difference manifest, let us introduce the
cylindrical coordinates $(z,\rho)$ as follows:

\beq
z\,=\,-r\,{\rm cos }\,\theta\,\,,
\,\,\,\,\,\,\,\,\,\,\,\,\,\,\,\,\,\,\,\,\,\,\,\,\,\,\,\,
\rho\,=\,r\,{\rm sin }\,\theta\,\,.
\label{snueve}
\eeq
The behaviour of the coordinate $\rho$ in the region in which 
$r\rightarrow\infty$ is of special interest to interpret the
asymptotic behaviour of the brane. It is not difficult to find how
$\rho$ depends on $\theta$ for the solution (\ref{ssiete}):

\beq
\rho\,=\,A\,
\Bigl(\,{\rm tan}\,{\theta\over 2}\,\Bigr)^{{1\over 1-2\nu}}\,\,.
\label{setenta}
\eeq
Clearly, when $\nu<1/2$, $\rho\rightarrow\infty$ for 
$\theta\rightarrow\pi$, whereas if $\nu>1/2$ the coordinate $\rho$
diverges for  $\theta\approx 0$. This implies that our solution does
not behave as a tube
\footnote{A similar calculation for $p<6$ gives $\rho\rightarrow{\rm
constant}$ at the spikes.}. This fact is more explicit if we rewrite
our solution (\ref{ssiete}) as a function $z=z(\rho)$. After a short
calculation, we get:

\beq
z\,=\,{\rho^{2\nu}\over 2\,A^{1-2\nu}}\,\,
\Bigl(\,\rho^{2(1-2\nu)}\,-\,A^{2(1-2\nu)}\,\Bigr)\,\,.
\label{stuno}
\eeq
When $\rho\rightarrow\infty$, the function $z(\rho)$ behaves
as:
\begin{figure}
\centerline{\hskip -.8in \epsffile{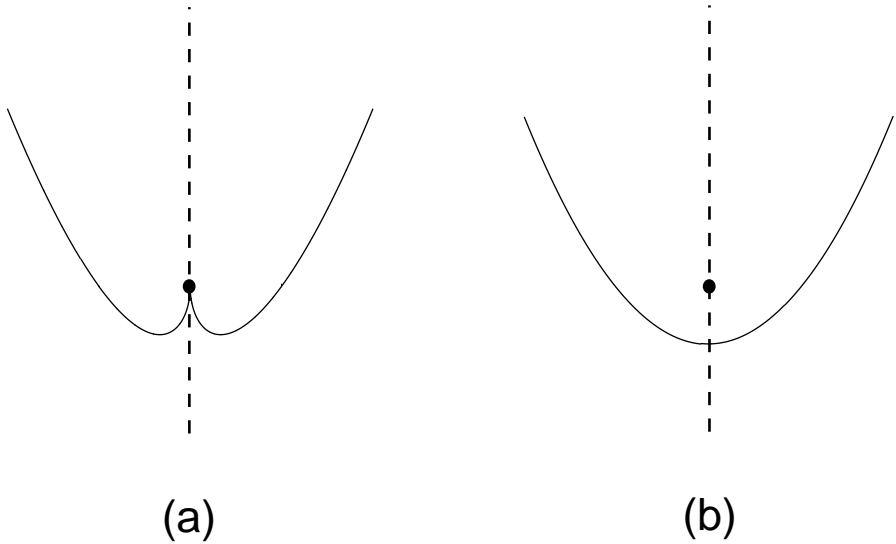}}
\caption{Representation of the D2 embedding for $\nu=1/4$ (a) and 
$\nu=0$ (b). The conventions are the same as in figure 2.}
\label{fig3}
\end{figure}

\beq
\lim_{\rho\rightarrow\infty}\,z(\,\rho\,)\,\sim\,
\cases{\rho^{2(1-\nu)}\,\,,&
            \,\,\,\,\,if $\nu\,<\,{1\over 2}\,\,,$\cr\cr
       -\rho^{2\nu}\,\,,&
            \,\,\,\,\,if $\nu\,>\,{1\over 2}$\,\,,\cr
}
\label{stdos}
\eeq
and thus, as expected, $z\rightarrow+\infty$ ($z\rightarrow-\infty$)
for $\nu<1/2$ ($\nu>1/2$). Moreover, eq. (\ref{stdos})  implies that
the asymptotic shape of the solution is that of a paraboloid
\footnote{ Strictly speaking, only for $\nu\,=\,0,1$ eq. (\ref{stuno})
represents a paraboloid.} rather than a tube. We have thus an ``upper
paraboloid" for $\nu<1/2$ and a ``lower paraboloid" for 
$\nu>1/2$. In figure 3 we have plotted the $p=6$ solution for
different values of $\nu$. Notice that for $\nu\,\not=\,0,1$ $z(\rho)$
has an extremum for some value of $\rho\not=0$, while for 
$\nu\,=\,0,1$ the extremum is located at $\rho=0$. This extremum is a
minimum (maximum) for  $\nu<1/2$ ($\nu>1/2$). It is interesting to
point out that, although the solution for $p=6$ differs from the ones
found for
$p\le 5$, the energy $U$ is still given by $|Z_s|$, where $Z_s$ is the
same as in eq. (\ref{stres}). 

It follows from our previous analysis that $\nu=1/2$ is a critical
point in the behaviour of the D6-D2 system. The BPS condition 
(\ref{scinco}) is not defined in this case and we have to come back to
the equation of motion (\ref{vseis}). Fortunately, we will be able to
find the general solution of the field equation in this case. The key
point in this respect is the observation that, for $p=6$ and
$\nu=1/2$, the near-horizon energy density does not depend on
$\theta$ explicitly. Indeed, a simple calculation shows that:

\beq
U\,=\,T_2\,\Omega_1\,R\,\,\int d\theta\,
\sqrt{r^2\,+\,r\,'^{\,2}}\,\,,
\,\,\,\,\,\,\,\,\,\,\,\,\,\,\,\,\,\,\,\,\,\,\,\,\
\,\,\,\,\,\,\,\,\,\,\,\,\,\,\,\,\,\,\,\,\,\,\,\,\
(\,\nu\,=\,{1\over 2}\,)\,\,.
\label{sttres}
\eeq
This $\theta$-independence  implies the ``conservation law":

\beq
r\,'\,{\partial U\over \partial r\,'}\,-\,U\,=\,
{\rm constant}\,\,,
\label{stcuatro}
\eeq
or, using the explicit form of $U$ given in eq. (\ref{sttres}):

\beq
{r^2\over \sqrt{r^2\,+\,r\,'\,^2}}\,=\,C\,\,,
\label{stcinco}
\eeq
where $C\ge 0$ is constant. It is not difficult to integrate the
first-order equation (\ref{stcinco}). The result is:

\begin{figure}
\centerline{\hskip -.8in \epsffile{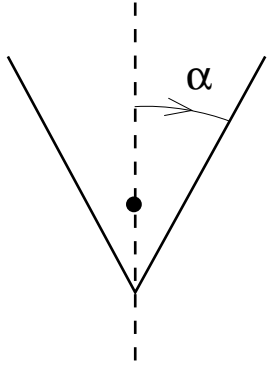}}
\caption{Solution of the equation of motion for $p=6$ and $\nu=1/2$
for the near-horizon metric (eq. (\ref{stseis})).}
\label{fig4}
\end{figure}

\beq
r(\,\theta\,)\,=\,{C\over {\rm sin }\,(\,\theta\,+\,\alpha\,)}\,\,,
\label{stseis}
\eeq
$\alpha$ being a new integration constant. The solution (\ref{stseis})
represents a cone with opening half-angle $\alpha$ and with its vertex
located at a distance $C/{\rm sin }\,\alpha$ from the origin (see figure
4). This fact is neatly shown if we rewrite (\ref{stseis}) in terms of
the cylindrical coordinates $(\rho, z)$ defined in eq. (\ref{snueve}). 
In these coordinates, eq. (\ref{stseis}) becomes:

\beq
\rho\,{\rm cos }\,\alpha\,-\,z\,{\rm sin }\,\alpha\,=\,C\,\,.
\label{stsiete}
\eeq
It is clear from figure 4 that $\alpha$ can take values in the range 
$-\pi/2\le\alpha\le\pi/2$. When $\alpha=0$ the cone degenerates in the
cylinder $\rho=C$, while for $\alpha\,=\,\pm\pi/2$ the solution 
(\ref{stsiete}) is the plane $z\,=\,\mp C$. Notice that eq. 
(\ref{stsiete})  is invariant under the transformation
$\alpha\rightarrow -\alpha$, $z\rightarrow -z$. Thus, for $\alpha>0$ 
($\alpha<0$) $z$ goes to $+\infty$ ($-\infty$) when the angle $\theta$
takes the value $\pi-\alpha$ ($-\alpha$). It is interesting to point
out that, when $\rho\rightarrow\infty$, 
$z$ diverges linearly with $\rho$, which is,
certainly, the  limiting case of the $\nu\not= 1/2$ situation (see eq.
(\ref{stdos})). This softer behaviour for $\nu= 1/2$ will be also
reflected in the solution for the full Dp-brane metric, as we shall
check in the next section, where we will find the solutions of
the BPS equation for the full Dp-brane brane metric (\ref{uno}). We
will check that for $p=6$ and $\nu=1/2$ this solution is equivalent, in
the near-horizon region, to our general solution (\ref{stsiete}) with 
$C=0$, \ie\ only those cones passing through the origin correspond to
BPS solutions of the asymptotically flat metric.

\setcounter{equation}{0}
\section{Asymptotically flat background}
\medskip

In this section we are going to study the solutions of the BPS
condition beyond the near-horizon approximation. Recall that the full
metric of the Dp-brane background is obtained by taking $a=1$ in the
harmonic function (\ref{dos}). The full Dp-brane metric is
asymptotically flat and, therefore, one expects that a brane placed
far away from the location of the background branes
is not bent by the gravitational field. We shall concentrate first in
the study of this asymptotic behaviour and, afterwards, we will consider
the complete solution. 

It is convenient for our purposes to rewrite the BPS condition
(\ref{tdos}) in the coordinates $(z,\rho)$ of eq. (\ref{snueve}). In
these coordinates, the D(8-p)-brane embedding is characterized by a
function  $z(\,\rho\,)$. The BPS equation gives  $d\,z/d\,\rho$ in
terms of  $(z,\rho)$. Actually, it is not difficult to relate 
$z(\,\rho\,)$ and its derivative to $r(\theta)$ and $r\,'(\theta)$. By
using only the relation between both coordinate systems one can verify
that:

\beq
{dz\over d\rho}\,=\,
{{\rm sin }\,\theta\,-\,{\rm cos }\,\theta\,\,\,{r\,'\over r}\over
{\rm cos }\,\theta\,+\,{\rm sin }\,\theta\,\,\,{r\,'\over r}}\,\,.
\label{stocho}
\eeq
By substituting the value of
the ratio $r\,'/ r$ given by eq. (\ref{tdos}) for $a=1$
on the right-hand side of this equation, one arrives
at the following first-order differential equation for the function 
$z(\,\rho\,)$:

\beq
{dz\over d\rho}=\,-
{[\,\rho^2\,+\,z^2]^{{7-p\over 2}}\over 
R^{7-p}\,+\,[\,\rho^2\,+\,z^2]^{{7-p\over 2}}}\,\,\,
{D_p\Bigr(\,{\rm arctan}\,(\,-\rho/z)\,\Bigr)\over
\rho^{7-p}}\,\,.
\label{stnueve}
\eeq
It is now elementary to evaluate the right-hand side of eq.
(\ref{stnueve}) in the asymptotic region in which
$\rho\rightarrow\infty$, $z/\rho\rightarrow 0$ 
(and $\theta\rightarrow\pi/2$). In this limit, eq. (\ref{stnueve})
takes the form:

\beq
{dz\over d\rho}\,\sim\,-\,{D_p(\,\pi/2\,)\over
\rho^{7-p}}\,+\,\cdots\,\,,
\label{ochenta}
\eeq
where we have only kept the first term in the expansion in powers of
$R/\rho$. Notice that $z\,'(\rho)\rightarrow 0$ as 
$\rho\rightarrow\infty$, according to our expectations. However, also
in this approach, the behaviour of the $p=6$ case differs from that
corresponding to $p<6$. Indeed, for $p=6$,  $z\,'(\rho)\sim \rho^{-1}$
and $z(\rho)$ diverges logarithmically in the asymptotic region while,
for $p<6$, $z(\rho)$ approaches a constant value as
$\rho\rightarrow\infty$. Let us specify further these two kinds of
behaviours. The coefficient multiplying the power of $\rho$ on the
right-hand side of eq. (\ref{ochenta}) is $D_p(\pi/2)$, which can be
computed from eq. (\ref{vdos}):

\beq
D_p(\pi/2)\,=\,-2\sqrt\pi\,{\Gamma\Bigl(\,{8-p\over 2}\Bigr)\over
\Gamma\Bigl(\,{7-p\over 2}\Bigr)}\,\,R^{7-p}\, 
(\,{1\over  2}\,-\,\nu\,)\,\,.
\label{ouno}
\eeq
Using this result we can integrate eq. (\ref{ochenta}). For $p<6$, we
get:

\beq
z\,(\,\rho\,)\,\sim\,z_{\infty}\,-\,(\,{1\over  2}\,-\,\nu\,)\,
\sqrt\pi\,\,
{\Gamma\Bigl(\,{6-p\over 2}\Bigr)\over
\Gamma\Bigl(\,{7-p\over 2}\Bigr)}\,\,R^{7-p}\, \,
{1\over \rho^{6-p}}\,+\cdots\,\,,
\,\,\,\,\,\,\,\,\,\,\,\,\,\,\,\,\,\,\,\,\,\,
(\,p\not=6\,)\,\,,
\label{odos}
\eeq
where $z_{\infty}$ is a constant representing the asymptotic value of
$z$. From eq.  (\ref{odos}) one concludes that the sign of
$z_{\infty}-z$ depends on the sign of 
${1\over  2}\,-\,\nu$. \ie\ if $\nu<1/2$ ($\nu>1/2$) the brane reaches
its asymptotic value of $z$ from below (above). If $p=6$ we get the
expected logarithmic dependence:

\beq
z\,(\,\rho\,)\,\sim\,(\,1\,-\,2\nu\,)\,R\,{\rm log}\,\rho
\,+\,\cdots\,\,.
\label{otres}
\eeq
Notice that when $\rho\rightarrow\infty$ in eq. (\ref{otres}),  
$z\rightarrow+\infty$ if $\nu<1/2$ while, on the contrary, 
$z\rightarrow-\infty$ for $\nu>1/2$, in agreement
with our near-horizon analysis of section 5. It is interesting to point
out that, although in this case $z$ diverges as
$\rho\rightarrow\infty$, 
$z/\rho$ still vanishes in this limit, as assumed in the derivation
of eq. (\ref{ochenta}). 

For $\nu=1/2$, the leading term in the asymptotic expansion vanishes
and we have to compute the next-to-leading term. This does not change
significantly the analysis of the $p<6$ case. However, for $p=6$, 
things change drastically when $\nu=1/2$. In fact, one can prove from
eq. (\ref{stnueve}) that, in this case, $z$ approaches a constant
value $z_{\infty}$ as $\rho\rightarrow\infty$ (and, actually, 
$z-z_{\infty}$ decreases as $\rho^{-1}$ for $\rho>>R$).

Let us now see how one can integrate exactly the BPS condition for the
full metric. The resulting solution must reproduce the asymptotic
behaviour we have just described. For $p=3$, eq. (\ref{stnueve}) was
integrated numerically in ref. \cite{CGS1}. As we shall show below, our
analytical solution agrees with these numerical results. 
It is more convenient to come back
to our original $(r,\theta)$ coordinates. The basic strategy we will
adopt to integrate the BPS equation is to write it in terms of the
same function $\Lambda_p(\,\theta\,)$ that we have already used to find
the near-horizon solutions. For illustrative purposes, we will do it for
a general value of the parameter $a$ of the harmonic function. After
using the definition (\ref{tnueve}), eq. (\ref{tdos}) can be put in
the form:

\beq
{r\,'\over r}\,=\,{a\,r^{7-p}\,(\,{\rm sin }\,\theta\,)^{8-p}\,+\,
R^{7-p}\,(\,{\rm sin }\,\theta\,)^{6-p}\,-\,
R^{7-p}\,\Lambda_p(\,\theta\,)\,{\rm cos }\,\theta\over
a\,r^{7-p}\,(\,{\rm sin }\,\theta\,)^{7-p}\,{\rm cos }\,\theta\,+\,
R^{7-p}\,\Lambda_p(\,\theta\,)\,{\rm sin }\,\theta}\,\,.
\label{ocuatro}
\eeq
Notice that eq. (\ref{ocuatro}) for $a=0$ is identical to eq.
(\ref{cuarenta}). Let us now isolate the terms of (\ref{ocuatro})
depending on $a$ in one of the sides of the equation. Doing this, it is
elementary to prove that eq. (\ref{ocuatro}) can be rewritten as:

\beq
a\,r^{7-p}\,(\,{\rm sin }\,\theta\,)^{7-p}\,
{d\over d\theta}\,\,(\,r\,{\rm cos }\,\theta\,)\,=\,
R^{7-p}\,\Bigl[\,r\,(\,{\rm sin }\,\theta\,)^{6-p}\,-\,
\Lambda_p(\,\theta\,)\,
{d\over d\theta}\,\,(\,r\,{\rm sin}\,\theta\,)\,
\Bigr]\,\,.
\label{ocinco}
\eeq
On the right-hand side of this expression we recognize a term
containing $({\rm sin }\,\theta\,)^{6-p}$, whose representation as a
derivative of the function $\Lambda_p(\,\theta\,)$ was crucial to
integrate the near-horizon BPS condition for  $p\not= 6$. Actually, 
assuming that $p\not= 6$ and using eq. (\ref{cdos}), our differential
equation becomes:

\bear
&&a\,r^{7-p}\,(\,{\rm sin }\,\theta\,)^{7-p}\,
{d\over d\theta}\,\,(\,r\,{\rm cos }\,\theta\,)\,=\,\rc\rc
&&={R^{7-p}\over 6-p}\,\Bigl[\,r\,{\rm sin }\,\theta\,\,
{d\over d\theta}\,\,\Lambda_p(\,\theta\,)-\,
(6-p)\,\Lambda_p(\,\theta\,)\,
{d\over d\theta}\,\,(\,r\,{\rm sin}\,\theta\,)\,
\Bigr]\,\,,
\,\,\,\,\,\,\,\,\,\,\,\,\,\,\,\,\,\,\,\,\,\,
(\,p\not=6\,)\,\,.\rc
\label{oseis}
\eear
Taking $a=0$ we recover the solutions (\ref{ctres}) and
(\ref{ciseis}). From now on we shall put  $a=1$ in all our expressions. 
In this case we can pass the 
$r^{7-p}\,(\,{\rm sin }\,\theta\,)^{7-p}$ factor in the left-hand side
 of eq.  (\ref{oseis}) to the right-hand side and, remarkably, this
equation can be written as:

\beq
{d\over d\theta}\,\,(\,r\,{\rm cos }\,\theta\,)\,=\,
{R^{7-p}\over 6-p}\,\,\,{d\over d\theta}\,\,
\Biggl[\,{\Lambda_p(\,\theta\,)\over 
(\,r\,{\rm sin }\,\theta\,)^{6-p}}\,\Biggr]\,\,,
\,\,\,\,\,\,\,\,\,\,\,\,\,\,\,\,\,\,\,\,\,\,\,\,\,\,\,\,\,\,\,\,
(\,p\not=6\,)\,\,.
\label{osiete}
\eeq
The integration of eq. (\ref{osiete}) is now immediate:

\beq
r\,{\rm cos }\,\theta\,=\,{R^{7-p}\over 6-p}\,\,
{\Lambda_p(\,\theta\,)\over 
(\,r\,{\rm sin }\,\theta\,)^{6-p}}
\,-\,z_{\infty}\,\,,
\,\,\,\,\,\,\,\,\,\,\,\,\,\,\,\,\,\,\,\,\,\,\,\,\,\,\,\,\,\,\,\,
(\,p\not=6\,)\,\,,
\label{oocho}
\eeq
where $z_{\infty}$ is a constant of integration which can be
identified with the asymptotic value of $z$ introduced in eq. 
(\ref{odos}). Actually, in terms of the coordinates $(z,\rho)$, our
solution (\ref{oocho}) can be written as:

\beq
z\,=\,z_{\infty}\,-\,{R^{7-p}\over 6-p}\,\,
\,\,{\Lambda_p
\Bigr(\,{\rm arctan}\,(\,-\rho/z)\,\Bigr)\over
\rho^{6-p}}\,\,,
\,\,\,\,\,\,\,\,\,\,\,\,\,\,\,\,\,\,\,\,\,\,\,\,\,\,\,\,\,\,\,\,
(\,p\not=6\,)\,\,.
\label{onueve}
\eeq
Eqs. (\ref{oocho}) and (\ref{onueve}) give the analytical solution of
the  BPS differential equation for $p\not=6$.  It is
easy to verify that eq. (\ref{onueve}) for $\rho\rightarrow\infty$ is
reduced to eq. (\ref{odos}). Moreover, by derivating 
both sides of eq.  (\ref{onueve}) with respect to
$\rho$, one can extract the value of
$dz/d\rho$. The result in terms of $(z,\rho)$ is precisely the one
written in (\ref{stnueve}), \ie\ the function $z(\rho)$ defined by 
(\ref{onueve}) is a solution of the BPS condition in cylindrical
coordinates.

\begin{figure}
\centerline{\epsffile{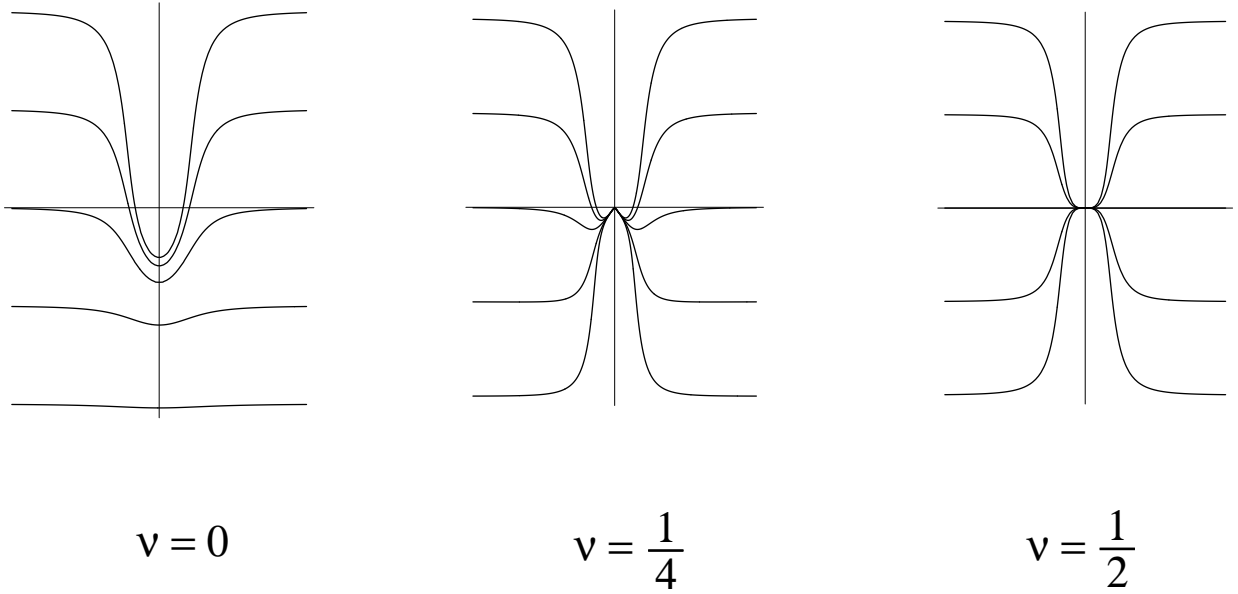}}
\caption{Solutions of the BPS differential equation for the
asymptotically flat metric (eq.~(\ref{onueve})) for $p=3$ and
several values of $\nu$ and $z_{\infty}$. }
\label{fig5}
\end{figure}

Let us now study  the analytical solution we have found
and, specially, how it interpolates between the near-horizon and
asymptotic regions. Notice, first of all,  that 
eqs.  (\ref{oocho})  and (\ref{onueve}) give $r(\theta)$ and $z(\rho)$
in implicit form. In particular, eq. (\ref{oocho}) determines
$r(\theta)$ as the root of a polynomial of degree $7-p$ which, for
$p\ge 3$, can be solved algebraically. However, the implicit equations 
(\ref{oocho})  and (\ref{onueve}) are much easier to handle and,
actually, can be used to plot the embeddings. In figure 5 we have
represented these embeddings for different values of $\nu$ and 
$z_{\infty}$. Moreover, by studying eqs. (\ref{oocho})  and
(\ref{onueve}) we can characterize the general features of our BPS
solutions. The detailed analysis of these solutions is performed in
appendix A. It follows from this analysis that our solution
(\ref{onueve}) coincides with the near-horizon embeddings in a region
close to the origin. Actually, as can be seen from the plots of figure
5, the integral of the BPS differential equation for the full metric
contains cylindrical regions which resemble closely  the tubes found
in the near-horizon analysis. It is not difficult to relate the
parameter $z_{\infty}$ of eq. (\ref{onueve}) to the constant $C$
appearing in the lower and upper tube solutions (eqs. (\ref{ctres}) and
(\ref{ciseis})). For large $|\,z_{\infty}\,|$ it can be seen that both
types of solutions are approximately equal if we are sufficiently near 
the origin and if the following relation:

\beq
C^{6-p}\,=\,{R^{7-p}\over (\,6\,-\,p\,)\,|\,z_{\infty}\,|}\,\,,
\label{noventa}
\eeq
is satisfied.

For $0<\nu<1$ the solution reaches the origin $r=0$ by means of a
tube, which is an upper or lower tube depending on the sign of 
$z_{\infty}$ (see figure 5). For $|\,z_{\infty}\,|\rightarrow\infty$
we will verify that this tubes can be regarded as bundles of
fundamental strings, exactly in the same way as in the near-horizon
case. 

The behaviour of the $\nu=0$ solution is different from that
we have just described for $0<\nu<1$\footnote{The solution for $\nu=1$
can be reduced to that for $\nu=0$ (see  eq. (\ref{apauno})).}. When 
$z_{\infty}\rightarrow-\infty$, the D(8-p)-brane is nearly flat. As we
increase $z_{\infty}$, an upper tube develops and for 
$z_{\infty}\rightarrow+\infty$ a bundle of $N$ fundamental strings,
connecting the Dp and D(8-p) branes, is created. As pointed out in
ref. \cite{CGS1}, this solution provides a concrete realization of the
so-called Hanany-Witten effect \cite{HW}.  

Let us now study the energy of our solutions. It is clear from the
plots of figure 5 that $r$ is not a single-valued function of
$\theta$. It is thus more convenient to use the $(z,\rho)$
coordinates in order to have a global description of the energy
functional. From eq. (\ref{docho}) it is straightforward to obtain 
the expression of $U$ as an integral over $\rho$. The result can be
put in the form:

\beq
U\,=\,T_{8-p}\,\Omega_{7-p}\,\,\int d\rho\,\,
\sqrt{\Bigl(\,\,\Delta_p\,-\,D_p\,{dz\over d\rho}\,\,\Bigr)^2\,+\,
\Bigl(\,\,D_p\,+\,\Delta_p\,{dz\over d\rho}\,\,\Bigr)^2}\,\,.
\label{nuno}
\eeq

In eq. (\ref{nuno}) $D_p$ depends on $\rho$ and $z(\rho)$ as in eq. 
(\ref{stnueve}) and $\Delta_p$ is the function defined in eq.
(\ref{vsiete}). It is important to stress that eq.~(\ref{nuno}) gives
the energy  for any embedding $z(\rho)$. Notice that on the right-hand
side of eq. (\ref{nuno}) we have the sum of two squares. The second of
these two terms vanishes when the BPS condition 
$dz/d\rho=-D_p/\Delta_p$ (see eq. (\ref{stnueve})) holds. It is
clear from these considerations that, if we define:

\beq
 X\,\equiv\,\Delta_p\,-\,D_p\,{dz\over d\rho}\,\,,
\label{extrauno}
\eeq
we obtain the following lower bound for the energy of any embedding:

\beq
U\,\ge\,T_{8-p}\,\Omega_{7-p}\,\,\int d\rho\,\,
\Bigl|\,X\,\Bigr|\,\,.
\label{extrados}
\eeq
The bound (\ref{extrados}) is saturated precisely when the embedding
satisfies the first-order BPS equation (\ref{stnueve}). A remarkable 
aspect of the function $X$ appearing in the right-hand side of eq. 
(\ref{extrados}) is that it can be put as a total derivative. Indeed,
one can verify that:

\beq
X\,=\,
{d\over d \rho}\,\,\Biggl[\,-z\,D_p\,+\,
\Bigl(\,{1\over 8-p}\,+\,
{R^{7-p}\over (\,\rho^2\,+\,z^2\,)^{{7-p\over 2}}}\,\Bigr)\,\,
\rho^{8-p}\,\,\Biggr]\,\,.
\label{extratres}
\eeq
Eq. (\ref{extratres}) is the analog in these coordinates of eq.
(\ref{tseis}). As it happened with eq. (\ref{tseis}), in order to
prove eq. (\ref{extratres}) one only has to use eq. (\ref{catorce})
and, therefore, (\ref{extratres}) is valid for any function $z(\rho)$. 
An important consequence of eq. (\ref{extratres}) is that the bound 
(\ref{extrados}) only depends on the boundary conditions of the 
embedding at infinity. Thus, one can say that the BPS embeddings are
those that minimize the energy for a given value of $z(\rho)$ at  
$\rho\rightarrow\infty$. From eq. (\ref{nuno}) it is very easy to obtain
the energy $U_{BPS}$ of one of such BPS solutions. The result one
arrives at is:

\beq
U_{BPS}\,=\,-\,T_{8-p}\,\Omega_{7-p}\,\,\int d\rho\,\,
\Bigl[\,{dz\over d\rho}\,+\, 
\Bigl(\,{dz\over d\rho}\,\Bigr)^{-1}\,\Bigr]\,\,D_p\,\,.
\label{ntres}
\eeq
Apart from being simple, eq. (\ref{ntres}) is specially suited for our
purposes. Notice, first of all, that the integrand on the right-hand
side of (\ref{ntres}) is always non-negative due to eq.
(\ref{stnueve}). Secondly, we can use (\ref{ntres}) to evaluate the
energy of a tubular portion of the brane. Indeed, 
for large $|z_{\infty}|$, 
$\Bigr|{dz\over d\rho}\Bigl|$ is large in one of these
tubes and the argument  of $D_p$
is almost constant and equal to $\pi$ (0) for an upper (lower) tube. 
Neglecting the term containing $\Bigl(\,{dz\over d\rho}\,\Bigr)^{-1}$
on the right-hand side of eq. (\ref{ntres}), and defining the length
of the tube as:

\beq
L_{tube}\,=\,\int_{tube}\,d\rho\,\Bigr|{dz\over d\rho}\Bigl|\,\,,
\label{ncuatro}
\eeq
we get the following values for the energy of the tubes:

\beq
U_{tube}\,=\,
\cases{-\,T_{8-p}\,\Omega_{7-p}\,D_p(\,\pi\,)\,L_{tube}\,\,,&
                   \,\,\,\,\,\,(upper tube)\,\,,\cr\cr
        T_{8-p}\,\Omega_{7-p}\,D_p(\,0\,)\,L_{tube}\,\,,&
                    \,\,\,\,\,\,(lower tube)\,\,.}
\label{ncinco}
\eeq

Taking into account eqs. (\ref{vtres}) and (\ref{vcinco}),  and using
eq. (\ref{vcuatro}), one can easily show from the result in eq. 
(\ref{ncinco}) that $U_{tube}$ coincides with $|Z_s|$, where the
values of $Z_s$ are given in eq. (\ref{stres}). This confirms our
interpretation of the tubes as bundles of $(1-\nu)N$ and $\nu N$
fundamental strings. 

It is not difficult to calculate the energy of the whole brane.
Actually, by means of eq.~(\ref{extratres}) we
can perform the integral appearing in the right-hand side of eq.
(\ref{ntres}). As we are calculating the energy of an infinite brane,
this integral is divergent. In order to regulate this divergence, let
us introduce a cutoff $\rho_c$, in such a way that the integral 
in eq. (\ref{ntres}) is
performed between $\rho=0$ and $\rho=\rho_c$. It is not difficult to
check from the properties of our solutions that the contribution of 
the lower limit $\rho=0$ is zero and, thus, only the value at the
cutoff contributes.  After this process, 
$U_{BPS}$ can be put as:

\beq
U_{BPS}\,=\,-T_{8-p}\,\Omega_{7-p}\,\,
\Biggl[\,z\,D_p\,-\,
\Bigl(\,{1\over 8-p}\,+\,
{R^{7-p}\over (\,\rho^2\,+\,z^2\,)^{{7-p\over 2}}}\,\Bigr)\,\,
\rho^{8-p}\,\,\Biggr]_{\rho\,=\,\rho_c}\,\,\,.
\label{nsiete}
\eeq
We have now to take $\rho_c\rightarrow\infty$ and, therefore, 
$z(\,\rho_c\,)\rightarrow z_{\infty}$. Notice that, in this limit, the
argument of the function $D_p$ is $\pi/2$. Using eq. (\ref{ouno}), 
$U_{BPS}$ can be put as a sum of two terms, namely:

\beq
U_{BPS}\,=\,\Bigl(\,{1\over 2}\,-\,\nu\,\Bigr)\,N\,T_f\,z_{\infty}\,+\,
T_{8-p}\,\Omega_{7-p}\,\,\Biggl[\,{\rho_c^{8-p}\over 8-p}\,+\,
R^{7-p}\,\rho_c\,\Biggr]\,\,.
\label{nocho}
\eeq
Notice that the first term on the right-hand side of eq. (\ref{nocho})
is finite and depends linearly on $z_{\infty}$, while the second term
diverges when $\rho_c\rightarrow\infty$ and is independent of 
$z_{\infty}$. According to  ref. \cite{Craps}, one can give the
following interpretation to this divergence. Let us consider a
D(8-p)-brane embedded in the metric  (\ref{uno}) along the plane $z=0$
or, equivalently, such that its 
worldvolume is determined by the equation  
$\theta=\pi/2$. Notice that in this configuration the brane is not bent
at all and, for this reason, it will be referred to as the ``ground
state" of the brane. Let
$g_{gs}$ be the induced metric on the worldvolume of the D(8-p)-brane
for this ground state configuration. Putting the worldvolume gauge
fields to zero  and substituting $g$ by $g_{gs}$ in eq. 
(\ref{seis}), we get the action of the ground state:

\beq
S_{gs}\,=\,-T_{8-p}\,\int d^{9-p}\,\xi\,
e^{-\tilde\phi}\,\,
\sqrt{-{\rm det}\,\,(\,g_{gs}\,)}\,\,.
\label{nnueve}
\eeq

The energy $E_{gs}$ of the ground state is obtained from $S_{gs}$ as:

\beq
E_{gs}\,=\,-{S_{gs}\over T}\,\,,
\label{cien}
\eeq
where $T=\int dt$. Using the metric given in eq. (\ref{uno})  and the
value of the dilaton field displayed in eq. (\ref{cinco}),
the calculation of  $E_{gs}$  is a simple exercise. By comparing the
result of this computation with the right-hand side of eq.
(\ref{nocho}), one discovers that $E_{gs}$ is equal to the divergent
contribution to $U_{BPS}$. Therefore, one can subtract $E_{gs}$ from 
$U_{BPS}$ and define a renormalized energy $U_{ren}$ as:
\beq
U_{ren}\,\equiv\,U_{BPS}\,-\,E_{gs}\,\,.
\label{ctuno}
\eeq
It follows from eq. (\ref{nocho}) that $U_{ren}$ is given by:

\beq
U_{ren}\,=\,
\Bigl(\,{1\over 2}\,-\,\nu\,\Bigr)\,N\,T_f\,z_{\infty}\,\,.
\label{ctdos}
\eeq
By derivating $U_{ren}$ with respect to $z_{\infty}$, we learn that
there is a net constant force acting on the D(8-p)-brane, which is
equal to $(\,{1\over 2}\,-\,\nu\,\Bigr)\,N\,T_f$, \ie\ equivalent to
the tension of $(\,{1\over 2}\,-\,\nu\,\Bigr)\,N$ fundamental strings.
In ref. \cite{CGS1} this force was interpreted as a consequence of the
fact that, due to the $p+2$-form flux captured by the D(8-p)-brane, the
latter is endowed with an effective charge equal to 
$(\,{1\over 2}\,-\,\nu\,\Bigr)\,N$ units.

Let us now integrate the BPS
differential equation for $p=6$. Using the value of $\Lambda_6(\theta)$
in eq. (\ref{ocinco}), we get (for $a=1$):

\beq
r\,{\rm sin }\,\theta\,\,
{d\over d\theta}\,\,(\,r\,{\rm cos}\,\theta\,)\,=\,R\,
\Bigl[\,r\,+\,(\,2\nu\,-\,1\,)\,
{d\over d\theta}\,\,(\,r\,{\rm sin}\,\theta\,)\,\Bigr]\,\,,
\,\,\,\,\,\,\,\,\,\,\,\,\,\,\,\,\,\,\,\,\,\,\,\,\,\,\,\,\,\,\,\,
(\,p=6\,)\,\,.
\label{cttres}
\eeq
This equation can be recast in the form:

\beq
{d\over d\theta}\,\,(\,r\,{\rm cos}\,\theta\,)\,=\,R\,
{d\over d\theta}\,\,\Biggl[\,{\rm log}\,\Bigl[\,
{\rm tan}\,\bigl(\,{\theta\over 2}\,\bigr)\,
(\,r\,{\rm sin}\,\theta\,)^{2\nu-1}\,\Bigr]\,\Biggr]
\,\,,\,\,\,\,\,\,\,\,\,\,\,\,\,\,\,\,\,\,\,\,\,\,\,\,\,\,\,\,\,\,\,\,
(\,p=6\,)\,\,,
\label{ctcuatro}
\eeq
and, therefore, its integration is immediate. It is more interesting
to write the result in cylindrical coordinates. After a short
calculation one gets:

\beq
\Bigl(\,z\,+\,\sqrt{\rho^2\,+\,z^2}\,\,\Bigr)\,
\rho^{2(\nu-1)}\,=\,e^{{k-z\over R}}\,\,
\,\,,\,\,\,\,\,\,\,\,\,\,\,\,\,\,\,\,\,\,\,\,\,\,\,\,\,\,\,\,\,\,\,\,
(\,p=6\,)\,\,,
\label{ctcinco}
\eeq
where $k$ is a constant of integration. It is easy to check that the
function $z(\rho)$ parametrized by eq. (\ref{ctcinco}) has the
asymptotic behaviour displayed in eq. (\ref{otres}). It follows that,
as expected, $z(\rho)$ does not reach a constant value for $p=6$ and 
$\nu\not=1/2$. The profiles of the $\nu \not=1/2$ curves are similar to
the near-horizon ones, plotted in figure 3. The only difference is
that, instead of the behaviour written in eq. (\ref{stdos}), 
$|\,z(\rho)\,|$ grows logarithmically as $\rho\rightarrow\infty$.
Moreover, similarly to what happens to the $p\le 5$ solution, 
eq. (\ref{ctcinco}) is invariant under the transformation:

\beq
\rho\rightarrow\rho\,,
\,\,\,\,\,\,\,\,\,\,\,\,\,
z\rightarrow -z\,,
\,\,\,\,\,\,\,\,\,\,\,\,\,
k\rightarrow -k\,,
\,\,\,\,\,\,\,\,\,\,\,\,\,
\nu\rightarrow 1 -\nu\,\,.
\label{ctseis}
\eeq

It is also easy to check that the function $z(\rho)$ defined by eq. 
(\ref{ctcinco})  coincides, in the near-horizon region, with the one
written down in eq. (\ref{stuno}) if the constants $A$ and $k$ are
identified as follows:

\beq
A^{2\nu-1}\,=\,e^{{k\over R}}\,\,,
\,\,\,\,\,\,\,\,\,\,\,\,\,\,\,\,\,\,\,\,\,\,\,\,\,\,\,\,\,\,\,\,\,\,
(\,\nu\,\not=\,{1\over 2}\,)\,\,.
\label{ctsiete}
\eeq

\begin{figure}
\centerline{\epsffile{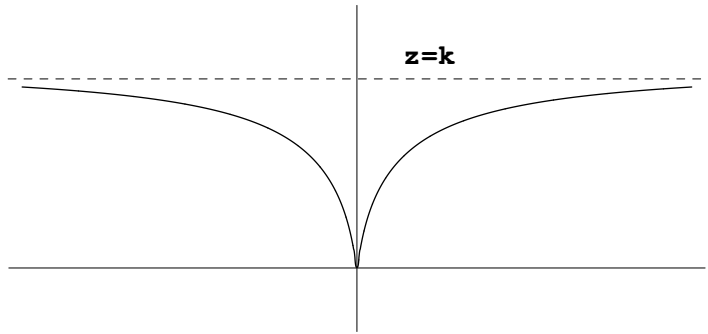}}
\caption{BPS embedding of the D2-brane in the full D6-brane geometry
for $\nu=1/2$. }
\label{fig6}
\end{figure}

For $\nu\,=\,1/2$ the behaviour of the solution changes drastically.
Indeed, in this case, one can solve eq. (\ref{ctcinco}) for $\rho$ and
get $\rho=\rho(z)$. The result is:

\beq
\rho\,=\,
{z\over {\rm sinh}\,{k-z\over R}}\,\,,
\,\,\,\,\,\,\,\,\,\,\,\,\,\,\,\,\,\,\,\,\,\,\,\,\,\,\,\,\,\,\,\,\,\,
(\,\nu \,=\,{1\over 2}\,,\,p=6\,)\,\,.
\label{ctocho}
\eeq
The function (\ref{ctocho}) has been represented graphically in figure
6. From this plot one notices that
$\lim_{\rho\rightarrow\infty}\,\,z(\rho)=k$. Actually, one can prove
using eq. (\ref{ctocho}) that for 
$k\ge 0$ ($k\le 0$) the variable $z$ takes values in the range 
$0\le z\le k$ ($k\le z\le 0$). Thus, in this ($p=6$, $\nu=1/2$) case,
the constant $k$ plays the same role as $z_{\infty}$ did for 
$p\le 5$. In particular, for $k=0$ the solution (\ref{ctocho}) is
equivalent to the equation $z=0$, which is precisely the same solution
we found for $p\le 5$, $\nu=1/2$ and $z_{\infty}=0$, 
\ie\ one obtains the ``ground state" solution in both cases.

Let us finally point out that, by linearizing eq. (\ref{ctocho}) in
$\rho$ and $z$, one gets the equation of a cone passing through the
origin and with half-angle equal to:

\beq
{\rm tan}\,\alpha\,=\,{1\over {\rm sinh}\,\,\Bigl(\,
{k\over R}\,\Bigr)}\,\,.
\label{ctnueve}
\eeq
Therefore, with the  identification (\ref{ctnueve}), we get a perfect
agreement with the near-horizon solutions (\ref{stsiete}) with $C=0$.

\setcounter{equation}{0}
\section{Summary and Conclusions}
\medskip

In this paper we have studied the embedding of a D(8-p)-brane in the
background geometry of a stack of coincident Dp-branes. This embedding
is governed by the worldvolume action of the D(8-p)-brane 
(eq. (\ref{uno})), which determines the equation of motion
(\ref{vseis}). By using a BPS argument we have found a bound for the
energy of the system such that those embeddings which saturate it 
are also a solution of the equation of motion. This equation of motion
(eq.~(\ref{tdos})) is a first-order differential equation which,
amazingly, can be solved analytically both in the near-horizon and
asymptotically flat geometries. 

The solutions of the BPS equations give the deformation of the
D(8-p)-brane under the influence of the gravitational and RR fields
created by the background branes. Generically, these solutions contain
tubes connecting the D(8-p)-brane to the background branes which,
after analyzing its energy, can be interpreted as bundles of
fundamental strings. From the point of view of the gauge theory defined
on the worldvolume of the background branes, these tubes represent
baryonic multiquark states. 

There are several topics which were not covered by our analysis which,
in our opinion, it would be interesting to consider in a future work.
Let us mention some of them. First of all, it would be interesting to
understand the relation between the  BPS  condition and
supersymmetry. One expects, following the 
line of ref. \cite{Imamura}, that the
BPS differential equation is precisely the requirement one must impose
to the brane embedding in order to preserve some fraction of the
space-time  supersymmetry.  An analysis of the supersymmetry algebra,
by applying the methods of ref. \cite{algebra}, could shed light on this
aspect of the problem.

Another topic which would be worth to explore is the use that our
exact results could have in the study of baryons in gauge theories
both in the supersymmetric and non-supersymmetric models.  In the
latter case we have to extend our results to the situation  in which the
background brane configuration is not extremal. 

Finally, it would be interesting to  find out if there exist BPS
conditions, similar to the ones found here, for brane embeddings in
more general brane geometries, such as the ones corresponding to the
intersection of several branes of different types. 

The final goal of these studies is twofold. On the one hand, we would
like to know what string theory can teach us about the
non-perturbative structure of gauge theories while, at the same time,
we would like to uncover aspects of a more complete formulation of
string theory.

\section{ Acknowledgments}
We are grateful to I. P. Ennes and P. M. Llatas for useful
discussions and a critical reading of the manuscript. This work was
supported in part by DGICYT under grant PB96-0960,  by CICYT under
grant  AEN96-1673 and by the European Union TMR grant
ERBFMRXCT960012.

%\newpage

\vskip 1cm                                               
{\Large{\bf APPENDIX A}}                                 
\vskip .5cm                                              
\renewcommand{\theequation}{\rm{A}.\arabic{equation}}  
\setcounter{equation}{0}  

In this appendix we are going to carry out an study of the solution
found in section 6 of the BPS differential equation for the
asymptotically flat metric for $p\le 5$ (eqs. (\ref{oocho}) and
(\ref{onueve})). 

The first thing we want to
point out in this respect is that, due to the property of
$\Lambda_p$ displayed in eq. (\ref{cisiete}),  our solution 
(\ref{onueve}) is invariant under the transformation:

\beq
\rho\rightarrow\rho\,,
\,\,\,\,\,\,\,\,\,\,\,\,\,
z\rightarrow -z\,,
\,\,\,\,\,\,\,\,\,\,\,\,\,
z_{\infty}\rightarrow -z_{\infty}\,,
\,\,\,\,\,\,\,\,\,\,\,\,\,
\nu\rightarrow 1 -\nu\,\,.
\label{apauno}
\eeq
Due to this invariance we can restrict ourselves to the case 
$0\le\nu\le 1/2$. The solutions outside this range of $\nu$ can be
obtained by performing a reflection with respect to the $z=0$ axis.
Therefore,  unless otherwise stated, we will assume in what follows
that   $\nu\le 1/2$. Moreover, it is more convenient for our purposes
to rewrite eq. (\ref{onueve}) in the equivalent form:

\beq
(\,z_{\infty}\,-\,z\,)\,\rho^{6-p}\,=\,
{R^{7-p}\over 6-p}\,\,\Lambda_p\,
\Bigr(\,{\rm arctan}\,(\,-\rho/ z)\,\Bigr)\,\,.
\label{apados}
\eeq
We will start our analysis of the solution by studying its cut with
the $\rho=0$ axis.  By looking
at eq. (\ref{apados}) it follows immediately that $\Lambda_p$ must
vanish for $\rho=0$. Thus, the angle $\theta$ at which the solution
reaches the $\rho=0$ axis must be $\theta\,=\,\theta_0$, $\theta_0$
being the same angle appearing in  the near-horizon solution (see eq. 
(\ref{cseis})). Following eq.  (\ref{snueve}), 
$\rho\,=\,r(\theta)\,{\rm sin}\,\theta$ and, therefore, 
if $\rho=0$ for $\theta=\theta_0$ one must have 
$r(\theta_0)\,{\rm sin}\,\theta_0\,=\,0$. There are two
possibilities to fulfill this equation. If, first of all, $\nu\not=0$,
the angle $\theta_0$ is non-vanishing, which means that, necessarily, 
$r(\theta_0)=0$, \ie\ the solution reaches the origin $r=0$ at an angle 
$\theta_0$. Notice that, as it should occur near  $r=0$,  this is
precisely what happens in the near-horizon solution. The second
possibility is  $\nu=0$. In this case $\theta_0=0$ and the
vanishing of $\rho$ does not require $r=0$. Again,  this is in agreement
with the near-horizon solution. Actually, the distance $r(0)$ at which
the 
$\nu=0$ solution cuts the $\rho=0$ axis can be determined, as in
section 4, by expanding $\Lambda_p(\theta)$ around $\theta_0=0$. Using
eq. (\ref{citres}) one gets that $r(0)$ must satisfy:

\beq
\bigl[\,r(0)\,\bigr]^{6-p}\,\,\,
\bigl[\,r(0)\,+\,z_{\infty}\,\bigr]\,=\,
{R^{7-p}\over 6-p}\,\,,
\,\,\,\,\,\,\,\,\,\,\,\,\,\,\,\,\,\,\,\,\,\,\,\,\,\,
(\,\nu\,=\,0\,)\,\,.
\label{apatres}
\eeq
As the right-hand side of this equation is positive, it follows that 
$r(0)\,>\,-z_{\infty}$. When this condition is satisfied, the
left-hand side of eq. (\ref{apatres}) is a monotonically increasing
function of $r(0)$ and, therefore, there exists a unique solution for
$r(0)$.

Let us now consider the cut with the $z=0$ axis. We shall denote
the corresponding value of the  $\rho$ coordinate by 
$\rho_0$, \ie\ 
$\rho_0\,=\,\rho(\,z=0\,)$. If $\rho_0\not= 0$, the  value
of the angle  $\theta$ for $z=0$ is $\theta=\pi/2$ and, after
evaluating the function $\Lambda_p$ for this value of $\theta$, eq. 
(\ref{apados}) gives:

\beq
\rho_0^{6-p}\,=\,2\sqrt{\pi}\,\,
{\Gamma\Bigl(\,{8-p\over 2}\Bigr)\over
\Gamma\Bigl(\,{7-p\over 2}\Bigr)}\,\,\,
({1\over 2}\,-\,\nu\,)\,\,\,{R^{7-p}\over z_{\infty}}\,\,.
\label{apacuatro}
\eeq

Notice that for $\nu\not=1/2$ the right-hand side of eq. 
(\ref{apacuatro}) only makes sense for $z_{\infty}>0$. Therefore, only
for $z_{\infty}>0$  the solution for $\nu\not=1/2$ cuts the $z=0$ axis
at a finite non-vanishing value of the coordinate $\rho$. For 
$\nu=1/2$ and $z_{\infty}\not=0$ eq. (\ref{apacuatro}) has no solution
for $\rho_0\not= 0$. All these features appear in the plots of figure
5.  

From eq. (\ref{apados}) one can extract the range of allowed values
of the coordinate $z$. Indeed, it follows from eq. (\ref{apados})
that the signs of $z_{\infty}-z$ and $\Lambda_p$ are the same. On the
other hand, we know that $\Lambda_p(\theta)$ is positive for 
$\theta>\theta_0$ and negative for $\theta<\theta_0$. Thus, when 
$\theta>\theta_0$ ($\theta<\theta_0$) one must have 
$z<z_{\infty}$ ($z>z_{\infty}$),  while for $\theta=\theta_0$ either
$\rho=0$ or else $z=z_{\infty}$. As 
$\theta_0\le\pi/2$ for $\nu\le 1/2$, one can easily see that these
results imply that 
$z\le 0$ for $z_{\infty}\le 0$,  whereas $z$ can be positive or
negative  if $z_{\infty}> 0$. 

It is easy to prove that $z$, as a function of $\rho$, must have a
unique extremum. For $\nu< 1/2$ this extremum is actually a minimum,
as we are going to verify soon. From eq. (\ref{stnueve}) it is clear
that $dz/d\rho$ vanishes if and only if $D_p$ is zero. Recall that the
Gauss' law (eq. (\ref{catorce})) implies that $D_p$ is a monotonically
decreasing function of $\theta$ (see figure 1). Moreover, it follows
from eqs. (\ref{vtres}) and (\ref{vcinco}) that $D_p(0)\ge 0$ and
$D_p(\pi)\le 0$. Thus, it must necessarily exist a unique value
$\theta_m$ of $\theta$ such that:

\beq
D_p(\,\theta_m\,)\,=\,0\,\,.
\label{apacinco}
\eeq
Clearly, at the point of the curve $z(\rho)$ at which $\theta=\theta_m$
the derivative $dz/d\rho$ is zero.  It is interesting to compare
$\theta_m$ with the angle
$\theta_0$ for which $\Lambda_p$ is zero. By substituting 
$\theta=\theta_0$ in the equation which relates 
$D_p(\,\theta\,)$ and $\Lambda_p(\,\theta\,)$ (eq. (\ref{tnueve})),
one gets:

\beq
D_p(\,\theta_0\,)\,=\,R^{7-p}\,
(\,{\rm sin }\,\theta_0\,)^{6-p}\,\,
{\rm cos }\,\theta_0\,\,.
\label{apaseis}
\eeq

For $\nu\le 1/2$ one has $\theta_0\,\le\,\pi/2$ and, thus, eq. 
(\ref{apaseis}) gives $D_p(\,\theta_0\,)\,\ge\,0$. The monotonic
character of $D_p(\,\theta\,)$ implies that 
$0\,\le\,\theta_0\,\le\,\theta_m\,\le\,\pi/2\,$. Notice that 
$\theta_0\,=\,\theta_m$ if $\nu=0$ ($\theta_0\,=\,\theta_m\,=\,0$) or 
$\nu=1/2$ ($\theta_0\,=\,\theta_m\,=\,\pi/2$).  The coordinate $z$ of
the extremum is 
$z_m\,=\,-r(\theta_m)\,{\rm cos }\,\theta_m$. As 
$\theta_m\le \pi/2$, one must have $z_m\le 0$. Moreover, since 
$\theta_m\ge \theta_0$, then $\Lambda_p (\theta_m)\ge 0$ and 
eq. (\ref{apados}) gives $z_m\le z_{\infty}$. By using the value of 
$\Lambda_p$ at $\theta=\theta_m$, which is:

\beq
\Lambda_p (\theta_m)\,=\,(\,{\rm sin }\,\theta_m\,)^{6-p}\,\,
{\rm cos }\,\theta_m\,\,,
\label{apasiete}
\eeq
(see eq. (\ref{tnueve})) one can obtain an expression which determines
$z_m$, namely:

\beq
|\,z_m\,|^{6-p}\,\,(\,z_{\infty}\,-\,z_m\,)\,=\,
{R^{7-p}\over 6-p}\,\,
\Bigr(\,{\rm cos }\,\theta_m\,\Bigr)^{7-p}\,\,.
\label{apaocho}
\eeq

It is not difficult to verify now that, when $\nu<1/2$, the extremum
at $\theta=\theta_m$ is a minimum. In order to prove  it, one must
evaluate  $d^2z/d\rho^2$. This can be done by derivating eq. 
(\ref{stnueve}). After putting 
$\rho\,=\rho_m\,=\,r(\theta_m)\,{\rm sin }\,\theta_m$, and using the
fact that $dz/d\rho$ vanishes for $\rho\,=\rho_m$, one arrives at:

\beq
{d^2\,z\over d\rho^2}\Big|_{\rho\,=\,\rho_m}
\,=\,-
{(\,7-\,p\,)\,\,R^{7-p}\over (\,\rho_m^{2}\,+\,z_m^2\,)\,\,
\Bigl[\,R^{7-p}\,+\,
(\,\rho_m^{2}\,+\,z_m^2\,)^{{7-p\over 2}}\,\Bigl]}\,\,\,z_m\,\,.
\label{apanueve}
\eeq
When $\nu<1/2$, one has $z_m<0$ and, as a consequence, the right-hand
side of eq. (\ref{apanueve}) is strictly positive. Therefore $z(\rho)$
has a minimum for $\rho\,=\,\rho_m$ as claimed. Notice that,  when
$\nu=1/2$, \ie\ $\theta_m=\pi/2$, the right-hand side of eq. 
(\ref{apaocho}) vanishes and, thus, $z_m=0$ (the value $z=z_{\infty}$
is reached asymptotically). 

Eq. (\ref{apaocho}) can be solved immediately if $z_{\infty}=0$.
Indeed, in this case, one has:

\beq
|\,z_m\,|^{7-p}\,=\,
{R^{7-p}\over 6-p}\,\,
\Bigr(\,{\rm cos }\,\theta_m\,\Bigr)^{7-p}\,\,,
\,\,\,\,\,\,\,\,\,\,\,\,\,\,\,\,\,\,
(\,z_{\infty}\,=\,0\,)\,\,.
\label{apadiez}
\eeq
For general values of $z_{\infty}$ we cannot give the general solution
of eq. (\ref{apaocho}).  However, one can extract valuable information
from this equation. An important point concerning eq. (\ref{apaocho}) 
is the fact that its right-hand side does not depend on $z_{\infty}$
(it only depends on $p$ and $\nu$) and, therefore, it remains constant
when we change the asymptotic value of the $z$ coordinate. This allows
us to study the behaviour of the minimum in different limiting
situations. Let us suppose, first of all, that
$z_{\infty}\,\rightarrow\,+\infty$ for
$\nu<1/2$. As $z_m<0$, the difference
$z_{\infty}-z_{m}\rightarrow\,+\infty$. In order to keep the
right-hand side of eq. (\ref{apaocho}) finite, one must have 
$z_m\rightarrow\,0^{-}$. Actually $z_m$ should behave as:

\beq
z_m\,\,\sim\,\,-{1\over (\,z_{\infty}\,)^{{1\over 6-p}}}\,\,,
\,\,\,\,\,\,\,\,\,\,\,\,\,\,\,\,\,\,
(\,z_{\infty}\,\rightarrow\,+\infty\,,\,\nu< {1\over 2}\,)\,\,.
\label{apaonce}
\eeq
On the other hand if
$z_{\infty}\,\rightarrow\,-\infty$, again for $\nu<1/2$, the only
possibility to maintain 
$|\,z_m\,|^{6-p}\,\,(\,z_{\infty}\,-\,z_m\,)$ constant is that
$z_{\infty}\,-\,z_m\rightarrow\,0^{+}$ (recall that 
$z_m\le z_{\infty}$ and, thus, $z_m\rightarrow\,0$ is impossible if 
$z_{\infty}\,\rightarrow\,-\infty$). The actual behaviour of 
$z_{\infty}\,-\,z_m$ in this limit is:

\beq
z_{\infty}\,-\,z_m
\,\,\sim\,\,{1\over (\,z_{\infty}\,)^{ 6-p}}\,\,,
\,\,\,\,\,\,\,\,\,\,\,\,\,\,\,\,\,\,
(\,z_{\infty}\,\rightarrow\,-\infty\,,\,\nu< {1\over 2}\,)\,\,.
\label{apadoce}
\eeq

The $\nu=0$ embedding presents some characteristics which make it
different from the $\nu\not= 0$ ones. We have already shown that the 
$\nu=0$  solution cuts the $\rho=0$ axis at a coordinate $z=-r(0)$,
with $r(0)$ determined by eq. (\ref{apatres}). Moreover, for 
$\nu=0$ the function $\Lambda_p$ on the right-hand side of 
eq. (\ref{apados}) is  non-negative for all $\rho\not=0$,
which implies that  $z\le z_{\infty}$ for these solutions. Recall
that, in this case, $z(\rho)$ has a minimum at $\rho=0$. For 
$z_{\infty}\,\rightarrow -\infty$, due to eq. (\ref{apadoce}),
$z(\rho)$ is approximately a constant function, \ie\ 
$z(\rho)\,\approx\,z_{\infty}$. If, on the contrary, 
$z_{\infty}\,\rightarrow +\infty$, an ``upper tube" connecting the
point $z=-r(0)\rightarrow 0$ (see eq. (\ref{apaonce})) 
and the asymptotic region is developed (see figure 5).

$\nu=1/2$ is also a special case. Indeed, for $\nu=1/2$ the angle
$\theta_0$ is equal to $\pi/2$ and, therefore, the function 
$\Lambda_p$ in eq. (\ref{apados}) takes positive (negative) values for 
$z>0$ ($z<0$). By studying the signs of both sides of eq. 
(\ref{apados}) one easily concludes that for $z_{\infty}>0$ one must
have $0\le z<z_{\infty}$, whereas for $z_{\infty}<0$ the coordinate
$z$ takes values on the range $z_{\infty}<z\le 0$ (see figure 5). When 
$z_{\infty}= 0$, the analysis of eq. (\ref{apados}) leads to the
conclusion that the solution is the hyperplane $z=0$.

It is also possible to determine the range of possible values that 
the coordinate  $\theta$ can take. Actually $\theta$ has generically an
extremal value, which we shall denote by $\theta_*$. The existence of 
$\theta_*$ can be proved by computing the derivative $d\theta/dr$ for
our solution. A short calculation  proves that this derivative
vanishes if the coordinate $z$ takes the value:

\beq
z_{*}\,=\,{6-p\over 7-p}\,\,z_{\infty}\,\,.
\label{apatrece}
\eeq
Alternatively, one could determine $z_{*}$ as the point at which the
denominator of the BPS condition (\ref{tdos}) vanishes. By computing
the second derivative $d^2\theta/dr^2$ at $\theta=\theta_*$, one can
check that $\theta_*$ is a minimum (maximum) if $z_{\infty}<0$ 
($z_{\infty}>0$)\footnote{For $\nu=0$ and $z_{\infty}<0$, the value
(\ref{apatrece}) is not reached (recall that $z\le z_{\infty}$ when
$\nu=0$). In this case the coordinate $\theta$ varies monotonically
along the embedding. }. It is also interesting to point out that for
$z=z_{*}$ the function ${\cal Z}$ of eq. (\ref{cinueve}) changes its
sign. This is consistent with the extremal nature of $\theta_*$.

\end{document}